\documentclass[twocolumn,numberedappendix,trackchanges]{aastex61}

\usepackage{natbib}
\usepackage{multirow}
\usepackage{gensymb} 
\usepackage[flushleft]{threeparttable} 
\usepackage{amsmath}
\usepackage{url}
\usepackage{multirow}
\usepackage{color}

\shorttitle{HST imaging of the brightest $z\sim8-9$ LBGs from UltraVISTA}
\shortauthors{Stefanon et al.}

\begin{document}

\title{HST imaging of the brightest $z\sim8-9$ galaxies from UltraVISTA: the extreme bright end of the UV luminosity function}

\author{Mauro Stefanon}
\affiliation{Leiden Observatory, Leiden University, NL-2300 RA Leiden, Netherlands}

\author{Ivo Labb\'e}
\affiliation{Leiden Observatory, Leiden University, NL-2300 RA Leiden, Netherlands}

\author{Rychard J. Bouwens}
\affiliation{Leiden Observatory, Leiden University, NL-2300 RA Leiden, Netherlands}

\author{Gabriel B. Brammer}
\affiliation{Space Telescope Science Institute, 3700 San Martin Drive, Baltimore, MD 21218, USA}

\author{Pascal Oesch}
\affiliation{Observatoire de Gen\`eve, 51 Ch. des Maillettes, 1290 Versoix, Switzerland}

\author{Marijn Franx}
\affiliation{Leiden Observatory, Leiden University, NL-2300 RA Leiden, Netherlands}

\author{Johan P. U. Fynbo}
\affiliation{Dark Cosmology Centre, Niels Bohr Institute, University of Copenhagen, Juliane Maries Vej 30, 2100 Copenhagen \O, Denmark}

\author{Bo Milvang-Jensen}
\affiliation{Dark Cosmology Centre, Niels Bohr Institute, University of Copenhagen, Juliane Maries Vej 30, 2100 Copenhagen \O, Denmark}

\author{Adam Muzzin}
\affiliation{York University, 4700 Keele Street, Toronto, ON, M3J 1P3, Canada}

\author{Garth D. Illingworth}
\affiliation{UCO/Lick Observatory, University of California, Santa Cruz, 1156 High St, Santa Cruz, CA 95064, USA}

\author{Olivier Le F\`evre}
\affiliation{Aix-Marseille Universit\'e, CNRS, LAM (Laboratoire d'Astrophysique de Marseille) UMR 7326, 13388 Marseille, France}

\author{Karina I. Caputi}
\affiliation{Kapteyn Astronomical Institute, University of Groningen, P.O. Box 800, 9700AV Groningen, The Netherlands}

\author{Benne W. Holwerda}
\affiliation{Department of Physics and Astronomy, University of Louisville, Louisville KY 40292 USA}

\author{Henry J. McCracken}
\affiliation{Institut dÕAstrophysique de Paris, 98bis Boulevard Arago, F-75014 Paris, France}

\author{Renske Smit}
\affiliation{Cavendish Laboratory, University of Cambridge, 19 JJ Thomson Avenue, Cambridge CB3 0HE, UK}
\affiliation{Kavli Institute of Cosmology c/o Institute of Astronomy, Madingley Road, Cambridge CB3 0HA, UK}

\author{Dan Magee}
\affiliation{UCO/Lick Observatory, University of California, Santa Cruz, 1156 High St, Santa Cruz, CA 95064, USA}

\email{Email: stefanon@strw.leidenuniv.nl}

\begin{abstract}

We report on the discovery of three especially bright candidate $z_\mathrm{phot} \gtrsim 8$ galaxies. Five sources were targeted for follow-up with \textit{HST}/WFC3, selected from a larger sample of 16 bright ($24.8 \lesssim H\lesssim25.5$~mag) candidate  $z\gtrsim 8$ LBGs identified over the 1.6 degrees$^2$ of the COSMOS/UltraVISTA field. These were identified as $Y$ and $J$ dropouts by leveraging the deep ($Y$-to-$K_\mathrm{S} \sim 25.3-24.8$~mag, $5\sigma$) NIR data from the UltraVISTA DR3 release, deep ground based optical imaging from the CFHTLS and Subaru Suprime Cam programs and \textit{Spitzer}/IRAC mosaics combining observations from the SMUVS and SPLASH programs. Through the refined spectral energy distributions, which now also include new HyperSuprime Cam $g, r, i, z,$ and $Y$ band data, we confirm that 3/5 galaxies have robust $z_\mathrm{phot}\sim8.0-8.7$, consistent with the initial selection. The remaining 2/5 galaxies have a nominal $z_\mathrm{phot}\sim2$.  However, if we use the HST data alone, these objects have increased probability of being at $z\sim9$. Furthermore, we measure mean UV continuum slopes $\beta=-1.91\pm0.26$ for the three $z\sim8-9$ galaxies, marginally bluer than similarly luminous $z\sim4-6$ in CANDELS but consistent with previous measurements of similarly luminous galaxies at $z \sim 7$.   The circularized effective radius for our brightest source is $0.9\pm0.2$ kpc, similar to previous measurements for a bright $z\sim11$ galaxy and bright $z\sim7$ galaxies.  Finally, enlarging our sample to include the six brightest $z\sim8$ LBGs identified over UltraVISTA (i.e., including three other sources from \citealt{labbe2017}) we estimate for the first time the volume density of galaxies at the extreme bright ($M_\mathrm{UV} \sim -22$~mag) end of the $z\sim 8$ UV LF. Despite this exceptional result, the still large statistical uncertainties  do not allow us to discriminate between a Schechter and a double power-law form.

\end{abstract}


\section{Introduction}

The study of the galaxy populations at the epoch of re-ionization has substantially improved in the last decade thanks to the exceptional sensitivity of the \textit{Hubble Space Telescope (HST)}/Wide Field Camera 3 (WFC3). Programs such as the Hubble Deep Field (\citealt{williams1996}), the Hubble Ultra-Deep and eXtreme-Deep Field (\citealt{beckwith2006, illingworth2013}), the Brightest of the Reionizing Galaxies (BoRG, \citealt{trenti2011}), the Hubble Frontier Fields (\citealt{lotz2017}) and the Cosmic Assembly Near-Infrared Deep Extragalactic Legacy Survey (CANDELS) / 3D-HST (\citealt{grogin2011, koekemoer2011,vandokkum2011,brammer2012, momcheva2016}) have enabled the identification of $\sim 1500$ candidate galaxies at $z>6$, $\sim200$ of which at $z\sim8-10$ (e.g., \citealt{oesch2013, bouwens2015, finkelstein2015a, bouwens2016}). These samples are characterised by $M_\mathrm{UV} \gtrsim -22$ (i.e., apparent magnitudes fainter than $\gtrsim 25.5$ at $z\sim8$), $\sim 1$~mag more luminous than the current determinations of the characteristic magnitude of the rest-frame ultraviolet (UV) luminosity functions (LFs). 

Given the steep faint-end slope of the UV LF at $z\gtrsim6$ (\citet{schechter1976} $\alpha\sim-2$; see e.g., \citealt{bouwens2011, mclure2013, schenker2013, duncan2014, bouwens2015, finkelstein2015a}), galaxies fainter than the characteristic luminosity  dominate the estimates of the star-formation rate density (e.g., \citealt{oesch2014,mcleod2015});  furthermore, under the hypothesis that the faint-end slope does not decrease at luminosities 3-4~mag fainter than current observational limits at $z\sim6-8$, their much higher (factors of $ 10^{2}-10^{4}$) volume density compared to the bright-end, has been proven sufficient for low-luminosity galaxies to complete the re-ionization by $z\sim6$ (e.g., \citealt{stark2016} and references therein). Nonetheless, given the correlation between the faint-end slope and the characteristic magnitude of the Schechter parameterization, commonly adopted to describe the shape of the LF at high redshift, the determination of the faint-end slope also benefits from improvements at the bright side. Furthermore, the identification of luminous galaxies at early epochs constitutes an important constraint to all models of galaxy formation and evolution. The recent spectroscopic confirmation of GN-z11, a luminous ($M_\mathrm{UV}=-22.1$~mag) galaxy at the record redshift of $z_\mathrm{grism} = 11.09_{-0.12}^{+0.08}$, has shown that its associated number density is higher than both the extrapolation to $z\sim10$ of the Schechter parameterization of the UV LFs and the model predictions (\citealt{oesch2016}), challenging our current understanding of galaxy formation and evolution.

The steep exponential decline at the bright end of the current UV LF determinations suggests that probing the LF at even brighter luminosities requires exploring areas of the order of a square degree in NIR bands to depths of $\sim25$~mag. Some progress in this direction has come from the BoRG and HIPPIES programs (\citealt{trenti2011, yan2011, bernard2016, calvi2016}), which have uncovered galaxies up to $z\sim8-10$ with $M_\mathrm{UV}\sim-22.5$~mag (e.g. \citealt{calvi2016}). A complementary approach comes from ground-based surveys, which allow us to extend the surveyed area to $\sim$1-100 deg$^2$, necessary to minimize the effects of cosmic variance in the systematic search for the brightest objects.

Recently, \citet{bowler2014, bowler2015, bowler2017} identified a sample of luminous galaxies ($-23\lesssim M_\mathrm{UV}\lesssim -22$ mag) at $z\sim6-7$ in the COSMOS/UltraVISTA field. Interestingly, the associated number densities are in excess of the \citet{schechter1976} form, suggesting that a double power-law could provide a better description at the bright end than the commonly assumed Schechter form. Partial confirmation to this comes from \citet{ono2017} who measured the bright end of the UV LF at $4 \lesssim z\lesssim7$ using data from the HyperSuprimeCam Survey (\citealt{aihara2017a, aihara2017b}). Analysis of this three-layered  dataset resulted in a sample of $\sim 600$ $z\sim6-7$ LBG galaxies ($\sim 70$ galaxies at $z\sim7$) with $M_\mathrm{UV}\lesssim -25 $~mag over $\sim100$ deg$^2$.  After carefully removing AGN contaminants, their UV LF measurements show an excess at the bright end of the UV LF compared to the Schechter parameterization from previous studies, although a double power law still over-predicts the brightest end.

 \begin{deluxetable}{rcc}
\tablecaption{Photometric depths of the adopted ground-based and \textit{Spitzer}/IRAC data sets, and corresponding average aperture corrections. \label{tab:depths}}
\tablehead{
\colhead{Filter} & \colhead{Aperture} & \colhead{Depth}  \\
\colhead{name} & \colhead{correction\tablenotemark{a}} & \colhead{$5 \sigma$\tablenotemark{b}} 
}
 \startdata
         CFHTLS $u^*$  & $    2.3 $ & $        26.7$ \\
              SSC $B$  & $    1.7 $ & $        27.4$ \\
              HSC $g$\tablenotemark{c}  & $    2.1 $ & $        26.7$ \\
           CFHTLS $g$  & $    2.2 $ & $        26.8$ \\
              SSC $V$  & $    2.2 $ & $        26.4$ \\
              HSC $r$\tablenotemark{c}  & $    1.6 $ & $        26.8$ \\
           CFHTLS $r$  & $    2.1 $ & $        26.4$ \\
            SSC $r^+$  & $    2.0 $ & $        26.6$ \\
           SSC  $i^+$  & $    1.9 $ & $         26.2$ \\
           CFHTLS $y$  & $    2.0 $ & $        26.1$ \\
           CFHTLS $i$  & $    2.0 $ & $        26.0$ \\
              HSC $i$\tablenotemark{c}  & $    1.7 $ & $        26.3$ \\
           CFHTLS $z$  & $    2.1 $ & $        25.2$ \\
              HSC $z$\tablenotemark{c}  & $    1.6 $ & $        25.9$ \\
           SSC  $z^+$  & $    2.3 $ & $        25.0$ \\
              HSC $y$\tablenotemark{c}  & $    2.1 $ & $        24.9$ \\
           UVISTA $Y$  & $    2.6 $ & $ 25.4 / 24.5 $ \\
          UVISTA $J$   & $    2.4 $ & $ 25.4 / 24.4 $ \\
           UVISTA $H$  & $    2.2 $ & $ 25.1 / 24.1 $ \\
UVISTA $K_\mathrm{S}$  & $    2.2 $ & $ 24.8 / 23.7 $ \\
       IRAC $3.6\mu$m  & $    5.3 $ & $ 24.9 / 24.5 $ \\
       IRAC $4.5\mu$m  & $    5.4 $ & $ 24.7 / 24.3 $ \\
       IRAC $5.8\mu$m  & $    8.4 $ & $        20.8$ \\
       IRAC $8.0\mu$m  & $   10.1 $ & $        20.6$ \\
 \enddata
 \tablenotetext{a}{Average multiplicative factors applied to estimate total fluxes.}
\tablenotetext{b}{Average depth over the full field corresponding to $5 \sigma$ flux dispersions in empty apertures of $1\farcs2$ diameter corrected to total using the average aperture correction. The two depths for UltraVISTA correspond to the ultradeep and deep stripes, respectively; the two depths for the \textit{Spitzer/}IRAC $3.6\mu$m and $4.5\mu$m bands correspond to the regions with SMUVS+SCOSMOS+SPLASH coverage (approximately overlapping with the ultradeep stripes) and SPLASH+SCOSMOS only ($\approx$ deep stripes).}
 \tablenotetext{c}{The HyperSuprimeCam data was not available during the initial selection of the sample; we included them in our subsequent analysis applying the same methods adopted for the rest of the ground and \textit{Spitzer/}IRAC mosaics.}
 \end{deluxetable}

\begin{deluxetable*}{cccccccccccccc}
\tablecaption{\textit{HST} observations we obtained over the bright $z\sim8-9$ candidates from \citet{labbe2017}. \label{tab:z8sample}}
\tablehead{
\colhead{ID} & \colhead{R.A.} & \colhead{Dec.} & \colhead{$H$} & \multicolumn{5}{c}{Exposure times} & \multicolumn{5}{c}{Photometric depths} \\
 & & & &  \colhead{$z_{098}$}  & \colhead{$Y_{105}$\tablenotemark{a}} & \colhead{$J_{125}$}  & \colhead{$JH_{140}$\tablenotemark{a}}& \colhead{$H_{160}$}  & \colhead{$z_{098}$}  & \colhead{$Y_{105}$\tablenotemark{a}} & \colhead{$J_{125}$}  & \colhead{$JH_{140}$\tablenotemark{a}}& \colhead{$H_{160}$} \\
 & \colhead{[J2000]} & \colhead{[J2000]} & \colhead{[mag]} &  \colhead{[sec]}  & \colhead{[sec]}  & \colhead{[sec]}  & \colhead{[sec]}  & \colhead{[sec]}  & \colhead{[mag]}  & \colhead{[mag]} & \colhead{[mag]} & \colhead{[mag]} & \colhead{[mag]} 
}
\startdata
UVISTA-Y-1 & 09:57:47.90 & 2:20:43.7 & $24.8$ & $1512$ & 1022 & $462$ & $1197$ & $412$ & $25.8$ & $25.6$ & $25.0$ & $25.5$ & $24.6$ \\
UVISTA-Y-5 & 10:00:31.89 & 1:57:50.2 & $24.9$ & $1512$ & $\cdots$ & $462$ & $\cdots$  & $412$ & $25.8$ & $\cdots$   & $25.0$ & $\cdots$  & $24.6$  \\
UVISTA-Y-6 & 10:00:12.51 & 2:03:00.5 & $25.3$ & $1512$ & $\cdots$ & $462$ & $\cdots$  & $412$ & $25.8$ & $\cdots$   &  $25.0$ & $\cdots$  & $24.6$  \\
UVISTA-J-1 & 10:02:25.48 & 2:29:13.6 & $24.6$ & $1512$ & $\cdots$ & $462$ & $\cdots$ & $412$ & $25.8$ &  $\cdots$  & $25.0$ & $\cdots$  & $24.6$ \\
UVISTA-J-2 & 09:59:07.19 & 1:56:54.0 & $24.4$ & $1512$ & $\cdots$ & $462$ & $\cdots$ & $412$ & $25.8$ & $\cdots$  & $25.0$ & $\cdots$  & $24.6$ \\
\enddata
\tablecomments{The limiting magnitudes refer to $5\sigma$ fluxes in apertures of $0\farcs6$ diameter corrected to total using the growth curve of point sources, consistent with the flux measurements in the WFC3 bands used in this work.}
\tablenotetext{a}{Fortuitously, observations in the $Y_\mathrm{105}$ and $JH_\mathrm{140}$ bands are available over over one $z\sim8-9$ candidate in our program as part of the separate \textit{HST} program SUSHI (PI: Nao Suzuki - PID: 14808).}
\end{deluxetable*}

In order to probe the bright-end of the UV LF at even higher redshift we leveraged the deep and ultradeep data of the third data release (DR3) of the UltraVISTA program (\citealt{mccracken2012}),  complemented by deep optical data from the CFHTLS \citep{erben2009, hildebrandt2009} and Subaru Suprime-Cam (\citealt{taniguchi2007}), and with deep IRAC mosaics we generated following \citet{labbe2015} which include observations from the SMUVS (PI: K. Caputi) and SPLASH (PI: P. Capak) programs.  Using LBG criteria we selected a sample of 16 bright ($H\sim24-25$~AB) galaxies at $z\sim8$ \citep{labbe2017}.

The primary question is if the bright sources identified from the ground-based selections exist or whether they are a population of lower-z interlopers.  Indeed, spectroscopic confirmation has recently been obtained for three UV-luminous ($M_\mathrm{UV}\sim -22$~mag) galaxies at $z\sim7.5-8.7$ with $H\sim25.1$~AB selected from CANDELS fields (\citealt{roberts-borsani2016}), one at $z\sim8.7$ (\citealt{zitrin2015}) and one at $z\sim11$ (GN-z11 - \citealt{oesch2016}).  Furthermore, the lower spatial resolution of ground-based observations, compared to HST data, can blend the signal from mergers or from physically unrelated objects and hence make them appear as single sources  (e.g., \citealt{bowler2017}), resulting in an over-estimate of the bright-end of the LF and an under-estimate of the LF at lower luminosities.  Photometric variability can be indicative of the presence of an AGN component or of a stellar or brown dwarf contaminant, which would introduce systematics or even contaminate the sample. Fluctuations in the signal induced by the random noise from the background can potentially conspire suppressing low signal-to-noise (S/N) signal at optical wavelengths and thus mimicking a high redshift solution.  Moreover, \citet{bowler2017} have shown that the electronics of the detectors can introduce image ghosts that can mimic high-redshift objects. While each of these effects are likely rare, we are looking for a small number of real high-redshift candidates in a large imaging dataset, and follow-up imaging is required to validate these candidates,  effectively eliminating many of these concerns.

We therefore selected five of the brightest candidate $z \sim 8-9$ LBGs from \citet{labbe2017} for targeted follow-up with \textit{HST}/WFC3 $z_{098}$, $J_{125}$ and $H_{160}$ bands (PI: R. Bouwens, PID: 14895) in order to attempt to confirm these sources and to better constrain their physical properties.   These candidates stood out for their unprecedented brightness ($24.5\lesssim H \lesssim 25.2$) and for their tantalizing plausible $z_\mathrm{phot} \sim 8.5-9.0$ solution, being detected in the UltraVISTA ultradeep stripes and non-detection in the deepest optical ground-based data.

This paper is devoted to presenting the results of the analysis of the HST data for the five candidate $z\sim8$ galaxies. In Sect. 2 we briefly describe the datasets and the selection criteria adopted for the assembly of the sample; in Sect. 3 we describe the \textit{HST} data and how the photometry was performed; the results are presented in Sect. 4; in Sect. 5 we discuss the results and in Sect. 6 we conclude.

Throughout this work, we use the following short-form to indicate \textit{HST}/WFC3 filters: WFC3/F098M $\rightarrow z_\mathrm{098}$; WFC3/F105W $\rightarrow Y_\mathrm{105}$; WFC3/F125W $\rightarrow J_\mathrm{125}$; WFC3/F140W $\rightarrow JH_\mathrm{140}$ and WFC3/F160W $\rightarrow H_\mathrm{160}$. We adopt a cosmology with $H_0=70$ km/s/Mpc, $\Omega_\Lambda=0.7$ and $\Omega_m=0.3$. All magnitudes are in the AB system.

\begin{figure*}
\includegraphics[width=18cm]{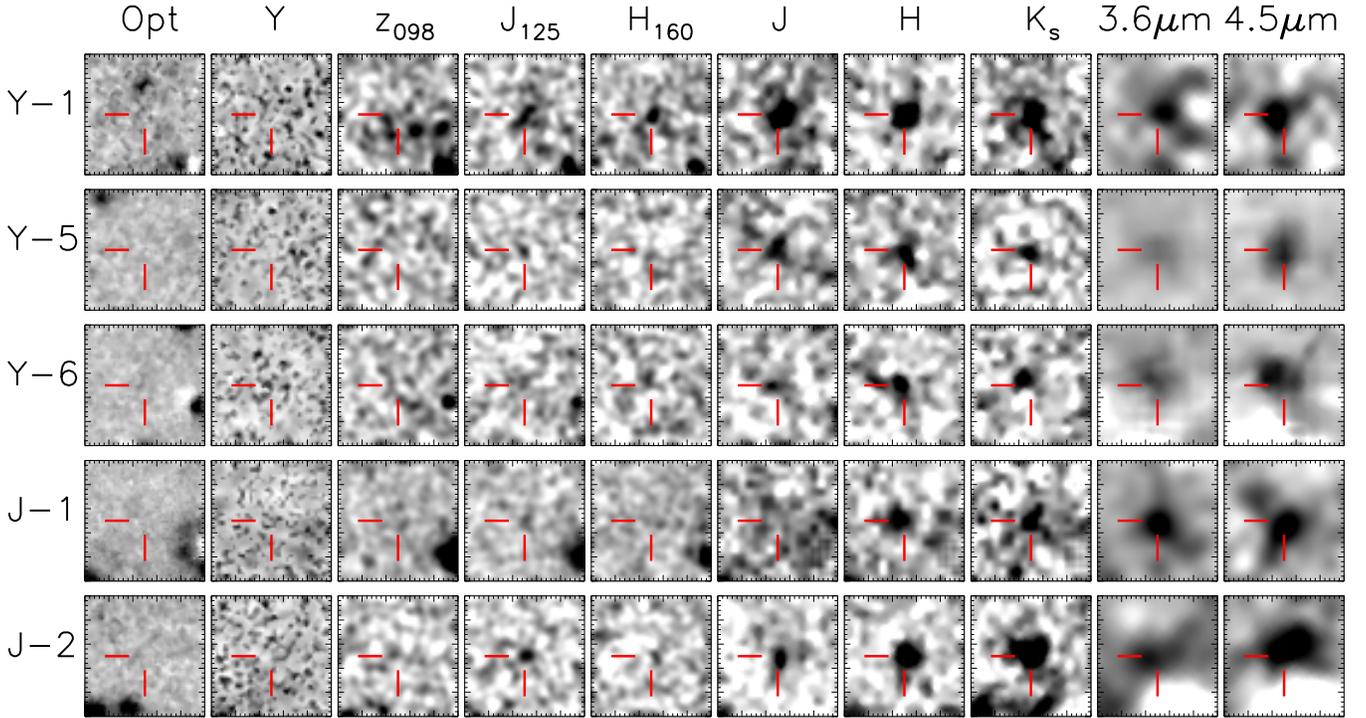}
\caption{Image stamps in inverted grey scale of the five bright candidate $z\gtrsim8$ LBGs in the stacked optical, stacked $Y$, \textit{HST}/WFC3 $z_\mathrm{098}$, $J_\mathrm{125}$, $H_\mathrm{160}$, UltraVISTA $J, H$ and $K_\mathrm{S}$ and \textit{Spitzer} IRAC $3.6\mu$m and $4.5\mu$m bands. Each row corresponds to a source, as labeled on the left, where we omitted the prefix UVISTA- from the object name for clarity. Each cutout is $5\farcs0  \times 5\farcs0$. The ground based and IRAC cutouts are shown after removing the neighbours. The cutouts in the \textit{HST}/WFC3 bands have been smoothed with a $\sigma=1.4$ pixels gaussian kernel. Postage stamp images of three other ultra-luminous ($M_\mathrm{UV} \lesssim -22$) $z\sim 8$ candidates are presented in Figure~\ref{fig:cuts_extra} from Appendix~\ref{app:extra}. \label{fig:cutouts}}
\end{figure*}

\section{Sample selection}
\label{sect:selection}
In this section we briefly summarize the dataset and the procedure followed to select the sample of candidate luminous $z\sim8$ sources in the COSMOS/UltraVISTA field. Full details are given in an accompaining paper (\citealt{labbe2017}). We give a concise summary below.

Our sample is based on the deep NIR imaging available over the COSMOS field (\citealt{scoville2007}) from the third release (DR3\footnote{\href{https://www.eso.org/sci/observing/phase3/data\_releases/uvista\_dr3.pdf}{https://www.eso.org/sci/observing/phase3/data\_releases/uvista\_dr3.pdf}}) of the UltraVISTA program (\citealt{mccracken2012}). This data release provides mosaics in the $Y, J, H$ and $K_\mathrm{S}$ broad bands together with a narrow band centered at $1.18\mu$m (NB118). The mosaics in the broad band filters are characterized by four \textit{ultradeep} stripes reaching $Y$-to-$K_\mathrm{S} \sim 24.8-25.3$~mag ($5\sigma, 1\farcs2$ aperture diameter corrected to total), alternating with four \textit{deep} stripes ($Y$-to-$K_\mathrm{S}\sim 23.7-24.4$~mag, $5\sigma, 1\farcs2$ aperture to total), for a total area of $\sim1.6$~degree$^2$. The UltraVISTA data were complemented by deep optical imaging from CFHT/Megacam in $g$, $r$, $i$ and  $z$ (\citealt{erben2009, hildebrandt2009}) from the Canada-France-Hawaii Legacy Survey (CFHTLS) and Subaru/Suprime-Cam (hereafter SSC) in $B_j$, $V_j$, $r^+$, $i^+$ and  $z$ bands (\citealt{taniguchi2007}). Full depth mosaics were constructed following  \citet{labbe2015} for the \textit{Spitzer}/IRAC $3.6\mu$m and $4.5\mu$m observations from S-COSMOS (\citealt{sanders2007}), the Spitzer Extended Deep Survey (SEDS; \citealt{ashby2013}), the Spitzer-Cosmic Assembly Near-Infrared Deep Extragalactic Survey (S-CANDELS, \citealt{ashby2015}), Spitzer Large Area Survey with Hyper-Suprime-Cam (SPLASH, PI: Capak), and the complete set of observations of the Spitzer Matching survey of the UltraVISTA ultra-deep Stripes (SMUVS, PI: Caputi - \citealt{caputi2017, ashby2017}).

Table \ref{tab:depths} lists the $5 \sigma$ depths of the adopted ground-based and \textit{Spitzer/}IRAC mosaics. They were measured as the standard deviation of fluxes in $\sim 4000$ empty apertures of $1\farcs2$ diameter, randomly scattered across the mosaic avoiding sources in the segmentation map. The values were finally multiplied by the average aperture corrections for each band (also reported in Table \ref{tab:depths}) to convert them into total fluxes, mimicking the procedures adopted for the flux measurements. While the exposure times across the CFHTLS, SSC and HSC mosaics are fairly uniform, the UltraVISTA and IRAC $3.6\mu$m and $4.5\mu$m mosaics are roughly characterized by a bi-modal depth. In these bands we therefore computed two different depths, restricting the random locations to regions representative of either one of the two typical depths. Our depth measurements are $\approx0.5-0.8$ mag brighter than previous estimates (e.g., \citealt{bowler2014, skelton2014}).  One possible reason for this is the specific statistical estimators adopted for the measurement. For instance, \citet{bowler2014} compute the background noise using the median absolute deviation (MAD). For a normal distribution, MAD is a factor $\sim1.5$ lower than the standard deviation, thus corresponding to $\sim 0.4$~mag fainter estimates. Finally, to ensure basic consistency with the results of \citet{bowler2014}, we independently made use of the MAD estimator to measure $5 \sigma$ depths and recovered values within $0.1$~mag from those presented by \citet{bowler2014}\footnote{This test was performed on $1\farcs8$ empty apertures for full consistency with \citet{bowler2014}.}.

Our search was carried out on the whole 1.6 degree$^2$ of the UltraVISTA field. The mosaics of the optical and ground-based NIR bands were PSF homogenized to the UltraVISTA $J$ band, so that the flux curve of growth for a point source would be the same across all bands. Fluxes from these bands were extracted using SExtractor (\citealt{bertin1996}) in dual mode.  Source detection was performed on the square root of the $\chi^2$ image (\citealt{szalay1999}) built using the UltraVISTA $J, H$ and $K_\mathrm{S}$ band science and rms-map mosaics. Total fluxes were computed from $1\farcs2$-diameter apertures and applying a correction based on the point-spread function (PSF) and brightness profile of each individual object.

Flux measurement for the \textit{Spitzer}/IRAC bands was performed with the \texttt{mophongo} software (\citealt{labbe2006, labbe2010a, labbe2010b, labbe2013, labbe2015});  briefly, the procedure consists in reconstructing the light profile of the objects in the same field of the source under consideration, using as a prior the morphological information from a higher resolution image. Successively, all the neighbouring objects within a radius of $15\farcs0$ from the source under analysis are removed using the positional and morphological information from the high resolution image and a careful reconstruction of the convolution kernel (see also e.g., \citealt{fernandez-soto1999,laidler2007,merlin2015}). Finally, aperture photometry is performed on the neighbour-cleaned source. For this work, we adopted an aperture of $1\farcs2$ diameter. The model profile of the individual sources is finally used to  correct the aperture fluxes for missing light outside the aperture. Specifically, this correction to \emph{total} flux is performed irrespective of detections or non-detections/negative flux measurements.

We note here that the use of morphological information and the kernel reconstruction operated by \texttt{mophongo} (similarly to other codes based on template fitting) renders unnecessary matching the images to the broadest PSF in the sample prior to extracting the flux densities, further reducing potential contaminations from neighbouring sources.

The sample of candidate galaxies at $z\gtrsim8$ was selected applying Lyman-Break criteria. Specifically, the following two color criteria were applied (\citealt{labbe2017}):
\begin{equation}
\label{eq:lbg}
Y - (J+H)/2 > 0.75 \lor (J-H)>0.8 
\end{equation}
for the Lyman break, and 
\begin{equation}
(H-K_\mathrm{S} < 0.7) \land (K_\mathrm{S}-[3.6] < 1.75 \lor H-[3.6] < 1.75)
\end{equation}
to reject sources with a red continuum red-ward of the J band, likely the result of a lower redshift dusty interloper. The symbols $\land$ and $\lor$ correspond to the logical AND and OR, respectively. Furthermore, sources showing $\ge2\sigma$ detection in any of the ground-based data blue-ward of the Lyman Break were removed from the sample. We note here that Eq.~(\ref{eq:lbg}) includes two different Lyman break criteria: the first one selects galaxies whose Lyman break enters the Y band, i.e., whose redshift is $\gtrsim 7.5$ and the second one selects galaxies whose Lyman break enters the J band, i.e, when the redshift is $\gtrsim 9.5$.

The sample was finally cleaned from potential brown dwarf contaminants. To this aim, we opted for not adopting SExtractor \texttt{class\_star} parameter because the classification becomes uncertain at low S/N (e.g., \citealt{bertin1996}). To overcome this, other star/galaxy separation criteria based on SExtractor have been developed (see e.g., \citealt{holwerda2014}). One of the most reliable is the effective radius vs magnitude (\citealt{ryan2011}). However, in order to  separate stars from galaxies, this method still requires to be applied to sources about $1.5$~mag above the photometric limit. Therefore, candidate brown dwarves were identified by fitting the observed SEDs with stellar templates from the SpecX prism library (\citealt{burgasser2014}) and from \citet{burrows2006} (which provide coverage up to $\sim15\mu$m for L and T dwarves) and excluding sources with $\chi^2$ from the stellar template set lower than from the galaxy templates.  The above selection criteria resulted in 16 candidate $z\gtrsim8$ galaxies brighter than $H=25.8$~mag. \\

Out of the 16 candidate galaxies at $z\gtrsim8$, we selected five (labelled UVISTA-Y-1, UVISTA-Y-5, UVISTA-Y-6, UVISTA-J-1 and UVISTA-J-2) with plausible $z_\mathrm{phot}\gtrsim8.5$ solutions, that stood out by their unprecedented brightness ($24.4\le H\le25.3$), which were detected in the UltraVISTA ultradeep stripes and with coverage from the deepest optical ground-based data in that region to be followed up with \textit{HST}/WFC3. Their positions and $H$-band fluxes are listed in Table \ref{tab:z8sample}.

\section{HST data and photometry}

\begin{deluxetable*}{lr@{ $\pm$}rr@{ $\pm$}rr@{ $\pm$}rr@{ $\pm$}rr@{ $\pm$}r}
\tablecaption{Total flux densities for the five candidate $z\gtrsim8$ LBGs over COSMOS/UltraVISTA targeted by our small \textit{HST} program. \label{tab:photometry}}
\tablehead{\colhead{Filter} & \twocolhead{UVISTA-Y-1} & \twocolhead{UVISTA-Y-5} & \twocolhead{UVISTA-Y-6} & \twocolhead{UVISTA-J-1} & \twocolhead{UVISTA-J-2} \\
& \twocolhead{[nJy]} & \twocolhead{[nJy]} & \twocolhead{[nJy]} & \twocolhead{[nJy]} & \twocolhead{[nJy]}
}
\startdata
CFHTLS $u^*$ &  \multicolumn2c{$\cdots$} & $    -13$ & $     17 $ & $      8$ & $     14 $ & $      4$ & $     15 $ & $     13$ & $     17 $ \\
SSC $B$ & $     -2$ & $      7 $ & $      4$ & $      9 $ & $    -11$ & $      9 $ & $     -4$ & $      8 $ & $      5$ & $     11 $ \\
HSC $g$ & $      8$ & $     15 $ & $    -10$ & $     20 $ & $      4$ & $     16 $ & $      1$ & $     15 $ & $      6$ & $     20 $ \\
CFHTLS $g$ &  \multicolumn2c{$\cdots$} & $      3$ & $     15 $ & $      1$ & $     13 $ & $     -8$ & $     10 $ & $     -4$ & $     17 $ \\
SSC $V$ & $    -12$ & $     17 $ & $     -1$ & $     24 $ & $      0$ & $     21 $ & $     -7$ & $     17 $ & $    -11$ & $     27 $ \\
HSC $r$ & $     -7$ & $     14 $ & $     -3$ & $     18 $ & $     11$ & $     15 $ & $      4$ & $     15 $ & $     10$ & $     17 $ \\
CFHTLS $r$ &  \multicolumn2c{$\cdots$} & $    -15$ & $     23 $ & $      8$ & $     21 $ & $    -22$ & $     17 $ & $      7$ & $     22 $ \\
SSC $r'$ & $      7$ & $     16 $ & $    -29$ & $     23 $ & $     20$ & $     19 $ & $     15$ & $     16 $ & $     -7$ & $     18 $ \\
CFHTLS $y$ &  \multicolumn2c{$\cdots$} & $    -11$ & $     28 $ & $     22$ & $     26 $ & $     16$ & $     23 $ & $    -26$ & $     28 $ \\
CFHTLS $i$ &  \multicolumn2c{$\cdots$} & $    -11$ & $     29 $ & $      5$ & $     29 $ & $    -28$ & $     24 $ & $    -29$ & $     29 $ \\
HSC $i$ & $     21$ & $     21 $ & $    -20$ & $     27 $ & $      1$ & $     23 $ & $     19$ & $     21 $ & $     23$ & $     27 $ \\
SSC $i^+$ & $    -9$ & $   22   $ & $    -36$ & $     26 $ & $      2$ & $     21 $ & $     -18$ & $     24 $ & $     -24$ & $     27 $ \\
CFHTLS $z$ &  \multicolumn2c{$\cdots$} & $      6$ & $     63 $ & $    -13$ & $     59 $ & $    -65$ & $     52 $ & $     32$ & $     62 $ \\
HSC $z$ & $      9$ & $     31 $ & $    -27$ & $     39 $ & $     17$ & $     33 $ & $     52$ & $     31 $ & $      8$ & $     37 $ \\
SSC $z'$ & $     51$ & $     64 $ & $    -51$ & $     93 $ & $     69$ & $     85 $ & $     70$ & $     64 $ & $    -39$ & $     83 $ \\
HSC $Y$ & $    -31$ & $     76 $ & $    -91$ & $     98 $ & $     89$ & $     80 $ & $     80$ & $     76 $ & $    -29$ & $     98 $ \\
$z_{098}$ & $     46$ & $     34 $ & $     -7$ & $     34 $ & $      9$ & $     34 $ & $     13$ & $     34 $ & $     22$ & $     34 $ \\
UVISTA $Y$ & $     18$ & $     48 $ & $    -42$ & $     68 $ & $     16$ & $     51 $ & $     67$ & $     55 $ & $     37$ & $     67 $ \\
$Y_{105}$ & $     92$ & $     41 $ &  \multicolumn2c{$\cdots$} &  \multicolumn2c{$\cdots$} &  \multicolumn2c{$\cdots$} &  \multicolumn2c{$\cdots$} \\
$J_{125}$ & $    279$ & $     70 $ & $    195$ & $     70 $ & $    172$ & $     70 $ & $    220$ & $     70 $ & $    304$ & $     70 $ \\
UVISTA $J$ & $    324$ & $     50 $ & $    235$ & $     66 $ & $    211$ & $     53 $ & $    125$ & $     50 $ & $    195$ & $     66 $ \\
$JH_{140}$ & $    303$ & $     44 $ &  \multicolumn2c{$\cdots$} &  \multicolumn2c{$\cdots$} &  \multicolumn2c{$\cdots$} &  \multicolumn2c{$\cdots$} \\
$H_{160}$ & $    511$ & $    107 $ & $    152$ & $    107 $ & $    272$ & $    107 $ & $    188$ & $    107 $ & $    257$ & $    107 $ \\
UVISTA $H$ & $    455$ & $     61 $ & $    393$ & $     86 $ & $    280$ & $     66 $ & $    510$ & $     70 $ & $    657$ & $     86 $ \\
UVISTA $K_\mathrm{S}$ & $    480$ & $     77 $ & $    321$ & $    102 $ & $    271$ & $     82 $ & $    602$ & $     95 $ & $    822$ & $    102 $ \\
IRAC $3.6\mu$m & $    623$ & $     85 $ & $    289$ & $     74 $ & $    434$ & $    106 $ & $   1162$ & $     90 $ & $    913$ & $    110 $ \\
IRAC $4.5\mu$m & $    931$ & $    109 $ & $    589$ & $     86 $ & $    598$ & $    130 $ & $   1277$ & $    102 $ & $   1204$ & $    130 $ \\
IRAC $5.8\mu$m & $  -2893$ & $   2568 $ & $  -1978$ & $   4831 $ & $   -643$ & $   3000 $ & $  -2209$ & $   2655 $ & $   8075$ & $   4602 $ \\
IRAC $8.0\mu$m & $   1423$ & $   3021 $ & $    499$ & $   6310 $ & $  -3325$ & $   3803 $ & $   1992$ & $   3771 $ & $   6734$ & $   5613 $ \\
\enddata
\tablecomments{Measurements for the ground-based and \textit{Spitzer/}IRAC bands are $1\farcs2$ aperture fluxes from \texttt{mophongo} corrected to total using the PSF and luminosity profile information; \textit{HST/}WFC3-band flux densities are measured in $0\farcs6$ apertures and converted to total using the PSF growth curves. Flux density measurements of three other ultra-luminous ($M_\mathrm{UV} \lesssim -22$) $z\sim 8$ candidates are presented in Table~\ref{tab:phot_extra} from Appendix~\ref{app:extra}.}
\end{deluxetable*}

The five bright candidate $z\sim8$ sources presented in this work benefit from \emph{HST}/WFC3 imaging obtained during the mid-cycle 24 (PI: R. Bouwens, PID: 14895). Observations were performed from March 27th, 2017 to March 29th, 2017. Table~\ref{tab:z8sample} summarizes the main observational parameters of the sample. Each source was observed for 1 orbit in total, subdivided as follows: $\sim1500$~s  ($\sim0.65$ orbits) in the F098M band ($z_\mathrm{098}$ hereafter), $\sim460$~s (0.18 orbits) in the F125W band ($J_\mathrm{125}$ hereafter), and $\sim 410$~s (0.17 orbits) in the F160W band ($H_\mathrm{160}$ hereafter). 

\begin{figure}
\hspace{-0.4cm}\includegraphics[width=9cm]{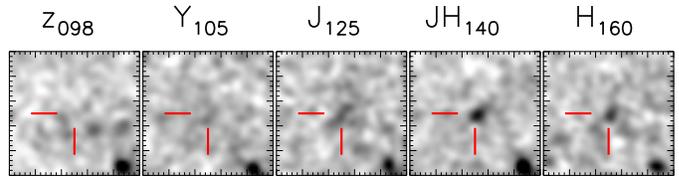}
\caption{Image stamps, in inverted grey scale, centered at the position of UVISTA-Y-1, in the five \textit{HST} bands available for this object, i.e. the three canonical bands targeted  by our \textit{HST} program (PI: R. Bouwens - $z_\mathrm{098}$, $J_\mathrm{125}$ and $H_\mathrm{160}$) plus $Y_\mathrm{105}$ and $JH_\mathrm{140}$ from the SUSHI program (PI: N. Suzuki).  The source is clearly detected in the $J_\mathrm{125}$, $JH_\mathrm{140}$ and $H_\mathrm{160}$ bands, while it is only slightly detected ($2.2\sigma$) in $Y_\mathrm{105}$. The cutouts in the \textit{HST}/WFC3 bands have been smoothed with a $\sigma=1.4$ pixels gaussian kernel.\label{fig:Y1_cutouts}}
\end{figure}

\begin{figure*}
\includegraphics[width=18cm]{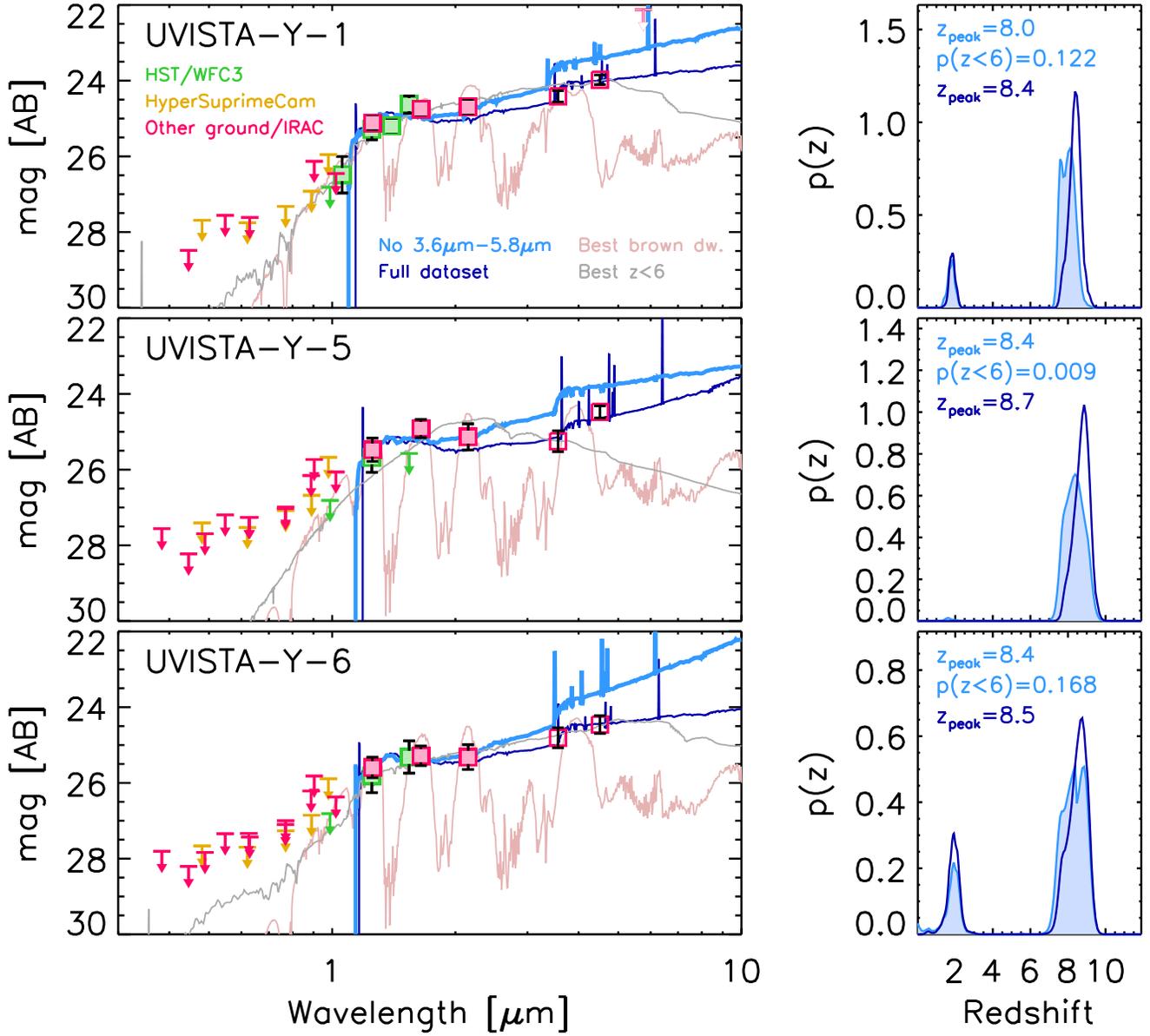}
\caption{{\bf Left panels:} SED for the three LBGs at $z\gtrsim 8$. The colored squares with black errorbars mark the photometric measurements, while arrows represent $2 \sigma$ upper limits. Open squares and arrows mark the IRAC $3.6\mu$m, $4.5\mu$m and $5.8\mu$m bands, not used for the measurement of the fiducial photometric redshift. Photometry in the \textit{HST} WFC3 bands is indicated by the green points and arrows; HyperSuprimeCam Survey data is represented in yellow. The fiducial best fit SED template from EAzY is indicated by the thick blue curve; the thin dark blue curve represents the best-fit SED when all bands are used for the photometric redshift measurement; the light brown curve presents the best-fitting brown-dwarf template, while the grey curve the solution obtained forcing the redshift to be $z<6$. {\bf Right panels:}  Redshift likelihood distributions ($p(z)$) for the three LBGs for the fiducial solution (blue) and for the solution obtained considering the full set of flux measurements. The label in the top-left corner indicates the estimated photometric redshifts. The $p(z)$ are peaked, with no or very low integrated probability for a secondary solution at lower redshifts. We caution the reader, however, that given the flux inconsistency between the $H_{160}$ and the UltraVISTA $H$ bands, the redshift estimate for  UVISTA-Y-5 may be less robust than that of the other two sources; further details are discussed in Sect. \ref{sect:SEDs} and Appendix \ref{app:flux}. The SEDs of three other ultra-luminous ($M_\mathrm{UV} \lesssim -22$) $z\sim 8$ candidates are presented in Figure~\ref{fig:extra} from Appendix~\ref{app:extra}. \label{fig:z9}}
\end{figure*}

\begin{figure*}
\includegraphics[width=18cm]{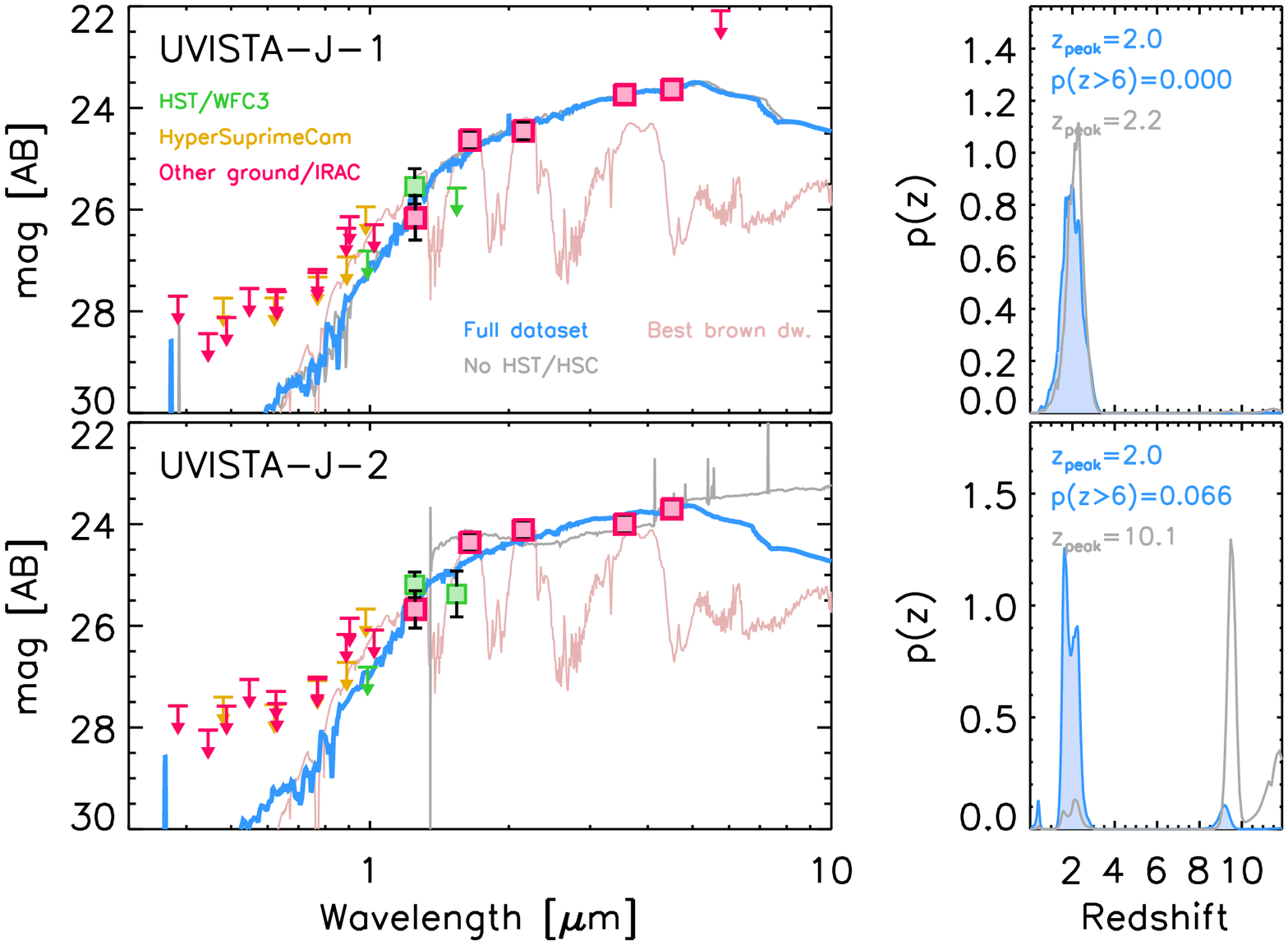}
\caption{SEDs and redshift likelihood distributions for the two galaxies with likely best solution at $z\sim2$. The result of excluding \textit{HST}/WFC3 and HSC measurements for the recovery of the photometric redshifts is marked by the grey curves.  Other plotting conventions as in Figure \ref{fig:z9}. We caution the reader, however, that given the flux inconsistency between the $H_{160}$ and the UltraVISTA $H$ bands, the redshift estimates for  these two sources may be less robust; further details are discussed in Sect. \ref{sect:SEDs} and Appendix \ref{app:flux}.\label{fig:z2}}
\end{figure*}

The field of UVISTA-Y-1 has also been observed by program 14808 (SUbaru Supernovae with Hubble Infrared - SUSHI - PI: Nao Suzuki) with $\sim 1000$~s integration time in the F105W band and $\sim1200$~s integration time in the F140W band, which we included in our analysis. Image stamps in all the five \textit{HST} bands centered at the position of UVISTA-Y-1 are shown in Figure \ref{fig:Y1_cutouts}.

The observations were processed using a customized version of multi-drizzle (\citealt{koekemoer2003}). For each object, the images in the three HST bands F098M, F125W and F160W were combined together into a \emph{red channel image}. Figure \ref{fig:cutouts} presents the image cutouts of the five objects in the three \textit{HST}/WFC3 bands together with ground based and \textit{Spitzer} IRAC bands.

Photometry of the HST bands was extracted using SExtractor in dual image mode, with the detection performed on the red channel image. Fluxes were measured in apertures $0\farcs6$ wide (diameter) in each band, and corrected for the flux excluded by the finite aperture using the PSF curve of growth. The typical aperture corrections were $\lesssim3\%$ across the WFC3 bands, minimizing potential systematic effects from the different PSFs.

Using the new HST data, we also reprocessed the flux measurements in all the ground-based optical and NIR bands and in the \textit{Spitzer}/IRAC bands. Fluxes were measured using the \texttt{mophongo} software in apertures $1\farcs2$ (diameter) and corrected to total using the light profile of each source. Remarkably, the optical data now include the mosaics from the HyperSuprimeCam Survey (HSC - \citealt{aihara2017a, aihara2017b}), not available at the time of the original selection of \citet{labbe2017}. This new program provides deep observations in the $g$, $r$, $i$, $z$ and $Y$ bands  ($5 \sigma$ depths of $26.6$, $26.7$, $26.2$, $25.8$ and $24.8$~mag, respectively). 

In Table \ref{tab:depths} we list the average multiplicative corrections applied to convert aperture fluxes into total. For ground-based data they range from $\sim 1.6$ to $\sim2.6$; for IRAC $3.6\mu$m and $4.5\mu$m data they are approximately $\sim5.4$, while for the two reddest IRAC bands they have values of $\sim 8-10$. The NIR and IRAC bands are characterized by large aperture corrections, which could introduce systematics in the estimates of the total fluxes. As a sanity check, we repeated the photometry with an  aperture of $1\farcs8$ diameter. The recovered total flux densities are on average within a few percent of the measurements based on the $1\farcs2$ apertures, and within $\sim15\%$ ($\sim 1 \sigma$) in the worst cases. We therefore considered the photometry obtained adopting the $1\farcs2$ aperture equally robust to that obtained with a larger aperture, but with a higher S/N.

Uncertainties associated to flux densities were computed differently depending on the dataset.

For \textit{HST/}WFC3 bands,  we estimated the noise associated to the background from the dispersion of values in 200 $0\farcs6$ apertures randomly placed across the image, free from sources according to the segmentation map, and repeating this process 20 times to increase the statistical significance. The final value was obtained applying the same aperture correction adopted for the estimate of the total fluxes. 

Uncertainties for ground-based and IRAC data were computed by \texttt{mophongo}. Briefly, the rms of the pixels in the residual image, obtained  subtracting all the detected objects from the science cutout, was computed for apertures of $1\farcs2$. As an additional step, the rms value was taken to be the maximum between the rms initially estimated by \texttt{mophongo} and the rms obtained from the empty apertures (see Sect. \ref{sect:selection} and Table \ref{tab:depths}). The systematic errors of kernel reconstruction were then added in quadrature and the result was scaled through the aperture correction.

The uncertainties resulting from this method are therefore not just the pure translation of the exposure map; specifically, the introduction of the systematic error from the kernel reconstruction and the scaling according to the aperture correction, which itself is, in general, different for different sources in a given band, makes the comparisons of uncertainties across different sources, in a given band, less immediate. Nonetheless, such method provides a robust and more comprehensive estimate of flux uncertainties.

One example for the above behaviour is given by the uncertainties in the UltraVISTA bands of UVISTA-Y-5 and UVISTA-Y-6. UVISTA-Y-6 lies at the border of one of the ultradeep stripes, while UVISTA-Y-5 is located in the middle of one of the ultradeep stripes. The ratio between the effective exposure time of UVISTA-Y-6 to that of UVISTA-Y-5 is $\sim0.76$, which would correspond to an increase of the rms background for UVISTA-Y-6 by a factor of $\sim1.14$. Instead, our analysis recovers flux uncertainties higher for UVISTA-Y-5 than for UVISTA-Y-6.  Inspection of the \texttt{mophongo} output showed that UVISTA-Y-5 is characterized by an rms background very similar to that of UVISTA-Y-6 and by a larger aperture correction ($\sim2.5$ versus $\sim2.0$, suggesting a more extended morphology for UVISTA-Y-5). The comparable values of the rms background for the two sources are likely the result of a larger value estimated by \texttt{mophongo} for the systematic uncertainty associated to the kernel reconstruction for UVISTA-Y-5.

As a further test, we repeated our analysis after replacing the uncertainties with the maximum uncertainty measured for each band across all the sources, and found results consistent with those from the main analysis, further supporting our error budget analysis.

The full set of measurements on ground- and space-based mosaics are presented in Table \ref{tab:photometry}. HST data were key to our work as they provided accurate positional and morphological priors for the \texttt{mophongo} photometry, enabling a more accurate neighbour subtraction. This, together with the additional information provided by the HSC survey (especially for UVISTA-Y-1 which lacks coverage from the CFHTLS survey), enabled a more accurate determination of the photometric redshifts and stellar population parameters for the galaxies in our sample.

\section{Results: Improved  spectral energy distributions and photometric redshifts}
\label{sect:SEDs}
The left panels of Figures \ref{fig:z9} and \ref{fig:z2} present the spectral energy distributions (SEDs) of the five sources studied in this work. The filled green squares and arrows mark the WFC3 measurements and $2\sigma$ upper limits, respectively.

In order to further assess their nature, we computed photometric redshifts running \texttt{EAzY} \citep{brammer2008} with the standard set of SED templates together with three old-and-dusty templates. Specifically, these templates correspond to a 2.5~Gyr, single burst, passively evolving, $Z_\odot$, \citet{bruzual2003} stellar population, further reddened with \citet{calzetti2000} $A_\mathrm{V}=2.0, 5.0, 8.0$~mag curves.

The input catalog consisted in the flux measurements listed in Table \ref{tab:z8sample} and \ref{tab:photometry}. One of the advantages of working with flux densities over magnitudes is that negative fluxes can be treated in a natural way, without any need to convert them into upper limits, thus preserving fidelity to observations.

Using the full set of bands, we find that UVISTA-Y-1, UVISTA-Y-5 and UVISTA-Y-6 have photometric redshifts $z_\mathrm{phot}=8.38^{+0.35}_{-0.43}, 8.74^{+0.45}_{-0.47}$ and $8.53^{+0.53}_{-0.80}$, respectively with $\chi^2=13.7, 10.6$ and $6.6$. The remaining two sources (UVISTA-J-1 and UVISTA-J-2) instead prefer solutions at $z_\mathrm{phot}\sim2$ ($\chi^2=18.5$ and $23.2$, respectively).

In Figure \ref{fig:z9} we also present the best-fitting brown dwarf SED template (light brown curve) and the best fit when we force the solution to be at $z<6$ (grey curve). Both these fits were obtained considering the full set of photometric points. Neither the brown dwarf nor the $z<6$ solutions do a better job at describing the observations compared to the $z\sim8$ best-fit template. Specifically, the brown dwarf template is inconsistent with observations in the IRAC $3.6\mu$m and $4.5\mu$m bands, and, for UVISTA-Y-1 also with the $JH_\mathrm{140}$ measurement. This is reflected by the poorer best-fit $\chi^2$'s, with $\chi^2 = 54.8, 58.5$ and $41.2$, respectively, for the 3 sources. Forcing the solution to be at $z<6$ results in $z_\mathrm{phot}\sim 2$. The best-fitting SED has a substantial contribution from an old/dusty component and provides a much better fit to the data than the brown-dwarf solution. However, it remains in tension with the data, resulting in $\chi^2$'s of $\chi^2 = 17.6, 25.3$ and $8.7$, respectively, for the three sources, i.e., $\Delta\chi^2 = 3.9, 14.7$, and $2.1$, respectively, worse than the $z\sim8$ fits. Similarly, the best-fit brown dwarf template for UVISTA-J-1 and UVISTA-J-2, displayed in Figure \ref{fig:z2}, are inconsistent with our measured fluxes in the IRAC $3.6\mu$m and $4.5\mu$m bands, where $\chi^2=61.3$ and $71.9$, respectively.

To ensure that our photometric redshift results are robust against potential errors and underestimates of the flux uncertainties for individual sources, we perturbed these by factors 1 to 1.5 randomly extracted from a uniform distribution. The new catalog was analysed following our standard procedure and the whole process was repeated 500 times. All of the recovered best-fit redshifts were within the $1 \sigma$ uncertainties of our nominal $z\sim8$ solutions.

We also looked at what happened to our photometric redshift solutions if the flux uncertainties were somewhat smaller than what we estimate, as for example we found in Sect.~\ref{sect:selection} (typical differences were factors of 1.5). We found photometric redshifts $z_\mathrm{phot} = 8.13^{+0.38}_{-0.42}, 8.57^{+0.47}_{-0.47}$ and $8.43^{+0.57}_{-0.89}$, for UVISTA-Y-1, UVISTA-Y-5 and UVISTA-Y6, respectively, with associated probabilities for the solution to be at $z>6$ of $p(z>6)=0.99, 0.99$ and $0.84$, respectively.

Limitations in our knowledge of the intrinsic SED shapes of $z\gtrsim 7$ galaxies (e.g., Balmer break amplitude, nebular emission lines equivalent width) make fits to the redder wavelength data more difficult, particularly in our attempts to derive accurate redshifts for the sources. During the SED fitting process, non-null colors from contiguous broad bands can be mis-interpreted as features which are not intrinsic to the source under analysis. For this reason, we repeated the photometric redshift measurements of the three $z\gtrsim8$ sources after excluding the IRAC $3.6\mu$m, $4.5\mu$m and $5.8\mu$m bands, as these are likely contaminated by strong emission lines and/or potentially contain the Balmer/4000 \AA~break. Both these properties are still poorly determined at these redshifts and any assumption about them could therefore introduce systematics in the redshift estimates. However, we still included the $8.0\mu$m data as it is likely not contaminated by strong nebular emission yet it provides constraints for the SED modelling. Indeed, the $K_\mathrm{S}-[4.5]$ color could be interpreted as the Balmer break, guiding the fit towards higher redshift solutions. These new measurements resulted in lower photometric redshifts: $z_\mathrm{phot}= 8.02^{+0.41}_{-0.49}, 8.39^{+0.60}_{-0.60}$ and $8.35^{+0.65}_{-0.81}$ for UVISTA-Y-1, UVISTA-Y-5 and UVISTA-Y-6, respectively. For this reason, we consider our fiducial photometric redshifts for UVISTA-Y-1, UVISTA-Y-5 and UVISTA-Y-6 those obtained \textit{without} the IRAC bands. We remark here, however, that the IRAC bands are nevertheless useful for our interpretation of these sources as they allow us to distinguish between genuine high-redshift sources and lower redshift interlopers.

The photometric redshift measurements for UVISTA-J-1 and UVISTA-J-2 were repeated after excluding the \textit{HST}/WFC3 and HSC flux measurements, to explore the possible reasons for the detected change in redshift. The redshift of UVISTA-J-2 obtained without the WFC3 and HSC bands is $z_\mathrm{phot}=10.1^{+1.4}_{-0.8}$, consistent with the initial selection. The new $H_{160}$ observations point to a much redder overall near-IR color (e.g., $H_{160}-[3.6]$) for the source, indicating that the $z<6$ solution is clearly the best one. For UVISTA-J-1, however, the photometric redshift we find does not sensibly change ($z_\mathrm{phot}=2.2^{+0.6}_{-0.4}$).  After further inspection, we conclude that the likely reason for this is a higher flux measurement in the $3.6\mu$m band we obtained using the new \textit{HST} dataset as morphological and positional prior which allowed for a more accurate subtraction of the neighbours, compared to the initial estimate obtained adopting the UltraVISTA bands. Both cases further stress the importance of high resolution imaging  from \textit{HST} in ascertaining the nature of candidate high-z sources. 

None of the five sources has a counterpart in the deep VLA catalogs of \citet{smolcic2017} nor in the X-Rays catalogs from \textit{XMM} and \textit{Chandra}  (\citealt{cappelluti2009} and \citealt{civano2016, marchesi2016}, respectively).  Finally, visual inspection of the MIPS $\mu$m mosaic from the S-COSMOS project \citep{sanders2007} did not show any evidence for the presence of sources at the nominal positions; we note, however, that a source is likely present $\sim 0\farcs8$ east of UVISTA-J-1.

The best-fit SEDs are shown as solid curves in the plot of the left-side panels of Figures \ref{fig:z9} and \ref{fig:z2}, while the right-side panels show the redshift likelihood generated by EAzY. In the following paragraphs we comment on the individual sources.

\textit{UVISTA-Y-1:} This source is undetected  ($<2\sigma$) in the HST/WFC3 $z_\mathrm{098}$ band, strengthening the evidence that this is a $z>7.5$ LBG.  The WFC3 photometry in the $J_\mathrm{125}$ and $H_\mathrm{160}$ is consistent at $1\sigma$ or better with that in the UltraVISTA $J$ and $H$ bands, respectively. This source also benefits from additional WFC3 coverage in the $Y_\mathrm{105}$ and $JH_\mathrm{140}$ bands from the SUSHI program (PI: N. Suzuki). The measurement in the $JH_\mathrm{140}$ band is consistent with the best-fit SED. The $p(z)$ is characterized by a solution at $ 8.02^{+0.41}_{-0.49}$, with a marginal secondary peak at $z\sim1.8$ ($p(z<6)=0.12$). The $Y_\mathrm{105}$ band shows a $2.2 \sigma$ detection, as expected if the source is at $z\sim8$.
 
\textit{UVISTA-Y-5:} The source is undetected ($<2\sigma$) in the HST/WFC3 $z_\mathrm{098}$ band, strongly favouring a $z>7.5$ solution for this source.  The WFC3 photometry in the $J_\mathrm{125}$ is consistent at $1\sigma$ with that in the UltraVISTA $J$ band.  However, the flux measurement in the $H_\mathrm{160}$ results in a $1.42\sigma$ detection only, and is consistent with the UltraVISTA H-band photometry at $\sim 3\sigma$ level. In Appendix \ref{app:flux} we analyse  more in detail the main effects that could explain the systematic differences observed in the $H_{160}$ and UltraVISTA $H$. Here we caution the reader that the observed discrepancy reduces our confidence on the high-redshift solution. The $p(z)$ is characterized by a peaked distribution at $ 8.39^{+0.60}_{-0.60}$, with a very marginal secondary peak at at $z\sim1.8$ ($p(z<6)=0.009$).

\textit{UVISTA-Y-6:} This source is undetected  ($<2\sigma$) in the HST/WFC3 $z_\mathrm{098}$ band, again favoring a $z>7.5$ solution.  The WFC3 photometry in the $J_\mathrm{125}$ and $H_\mathrm{160}$ is consistent at $1\sigma$ or better with that in the UltraVISTA $J$ and $H$ bands, respectively. The $p(z)$ shows a distribution with best fit solution $z_\mathrm{phot}= 8.35^{+0.65}_{-0.81}$ with a hint of secondary solution at $z<6$ ($p(z<6)=0.168$).

\textit{UVISTA-J-1:} The object is formally undetected in the $H_\mathrm{160}$ band ($<2\sigma$), making it consistent with the UltraVISTA $H$ band only at $\sim3.5\sigma$ level. We again refer the reader to Appendix \ref{app:flux} for a detailed discussion about possible origins of the observed difference, and flag this source because of the decreased confidence on the redshift determination. The photometry in the $J_\mathrm{125}$ band, instead, is consistent with that in the UltraVISTA $J$ band at $1 \sigma$ level. The lower-z solution is enforced by the fact that the source is detected in the HSC $z$ band at  $1.7\sigma$ level. The fiducial photometric redshift is $ 2.05^{+0.49}_{-0.46}$. The $p(z)$ shows a peaked distribution around $z\sim2$ with no further secondary peaks at higher redshifts.

\textit{UVISTA-J-2:} The HST WFC3 photometry in the $J_\mathrm{125}$ and $H_\mathrm{160}$ is consistent with that in the corresponding UltraVISTA bands at $\sim1.8-3\sigma$, respectively, with nominal redshift of $ 1.96^{+7.04}_{-0.33}$. Similarly to what done for UVISTA-Y-5 and UVISTA-J-1, in Appendix \ref{app:flux}  we analyse the main effects that could generate the observed difference between the $H_{160}$ and $H$ bands. Again, here we caution the reader that this discrepancy reduces our confidence on our redshift estimate. The large uncertainty associated to the upper limit makes it consistent with $z\sim9-9.5$; however, the $p(z)$ shows a pronounced peak at $z\sim2$ and a secondary peak at $z\sim10$, with a likelihood for the SED to be at $z>6$ of $p(z>6)=0.066$. For this reason we consider the fiducial redshift for this source to be the $z\sim2$ solution. \\

\section{Discussion}

\subsection{The brightest candidate LBGs at $z\gtrsim8$}

\begin{figure}
\hspace{-0.4cm}\includegraphics[width=9.5cm]{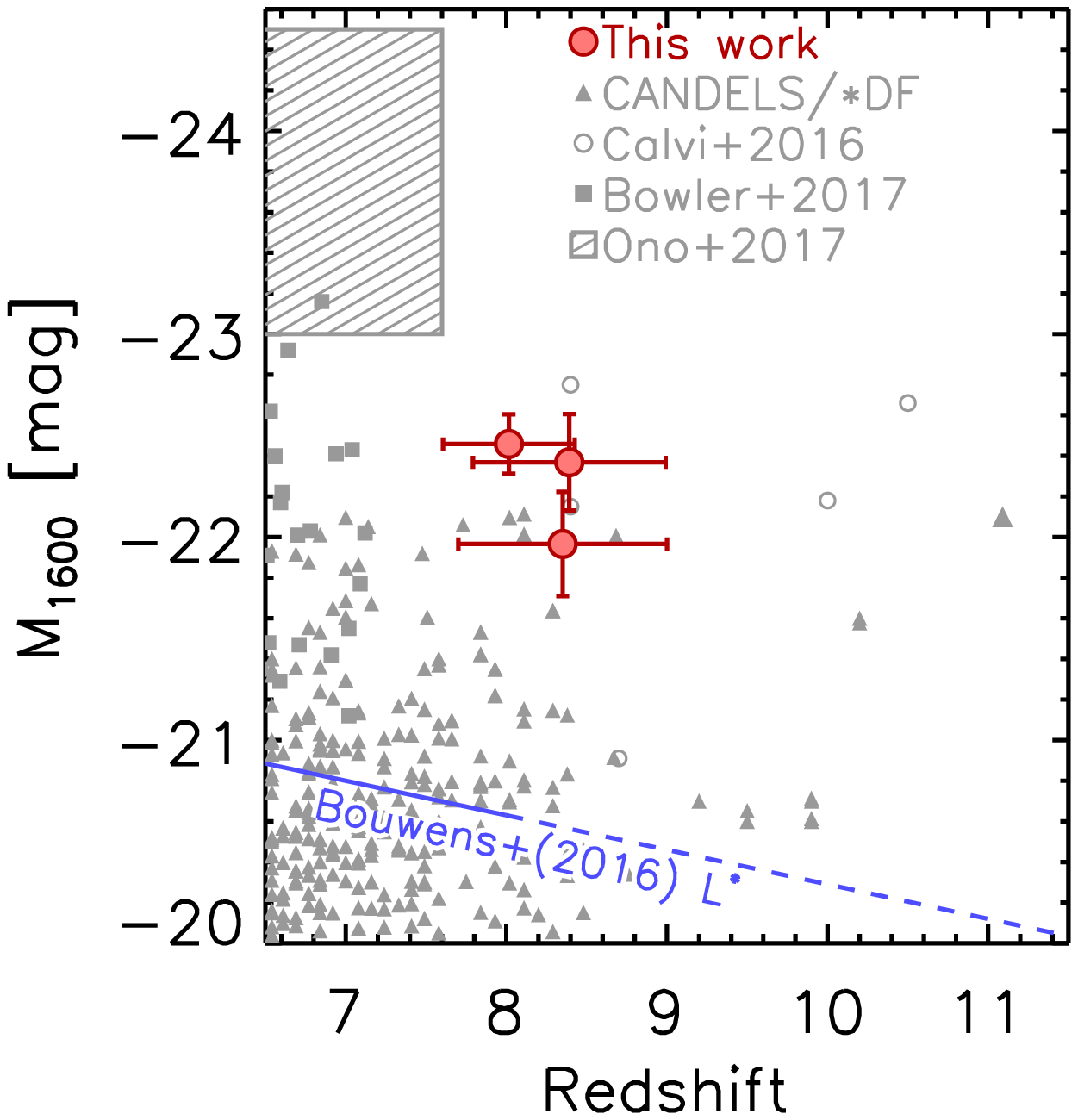}
\caption{Distribution with redshift of the absolute magnitude of the bright LBGs detected so far at $z\gtrsim7$.  We include measurements of \citet{oesch2014,bouwens2015, roberts-borsani2016,calvi2016,bowler2017} and \citet{ono2017} as indicated by the legend.  Our candidate LBGs are marked as red points with error bars and lie at the high luminosity end of all candidate $z\sim8$ LBGs to date.   Given the depth and wavelength coverage of the observations available for our sources, our bright sample arguably constitutes the brightest and most reliable sample of $z\sim8-9$ galaxies to date. \label{fig:zMUV}}
\end{figure}

Figure \ref{fig:zMUV} presents our sample of candidate $z\gtrsim8$ LBGs in the redshift-$M_\mathrm{UV}$ plane, together with recent LBG selections covering the bright-end of the UV LF at $z\gtrsim7$ of \citet{oesch2014} on XDF/HUDF,  \citet{bouwens2015}, \citet{roberts-borsani2016} and \citet{oesch2016} based on CANDELS data, \citet{calvi2016} from the BoRG program, \citet{bowler2017} from UltraVISTA and \citet{ono2017} from the HSC survey. We note however that the candidates of \citet{calvi2016} lack IRAC coverage and those of  \citet{ono2017} have measurements only at optical wavelengths from the HSC survey, resulting in their nature being more uncertain. The three $z\sim8$ galaxies reported on here constitute the brightest, most reliable $z\sim8-9$ galaxies discovered to the present. In order to put their luminosities in better context, in the same figure we also represent the evolutionary relation of the characteristic magnitude of the UV LF of \citet{bouwens2015} up to $z=8$ and its extrapolation to $z\sim10$. Our sample of luminous galaxies are $\sim1.8$~mag more luminous (a factor $\sim5.3\times$) than the estimated characteristic magnitude at $z\sim9$.

\begin{figure}
\hspace{-1cm}\includegraphics[width=10cm]{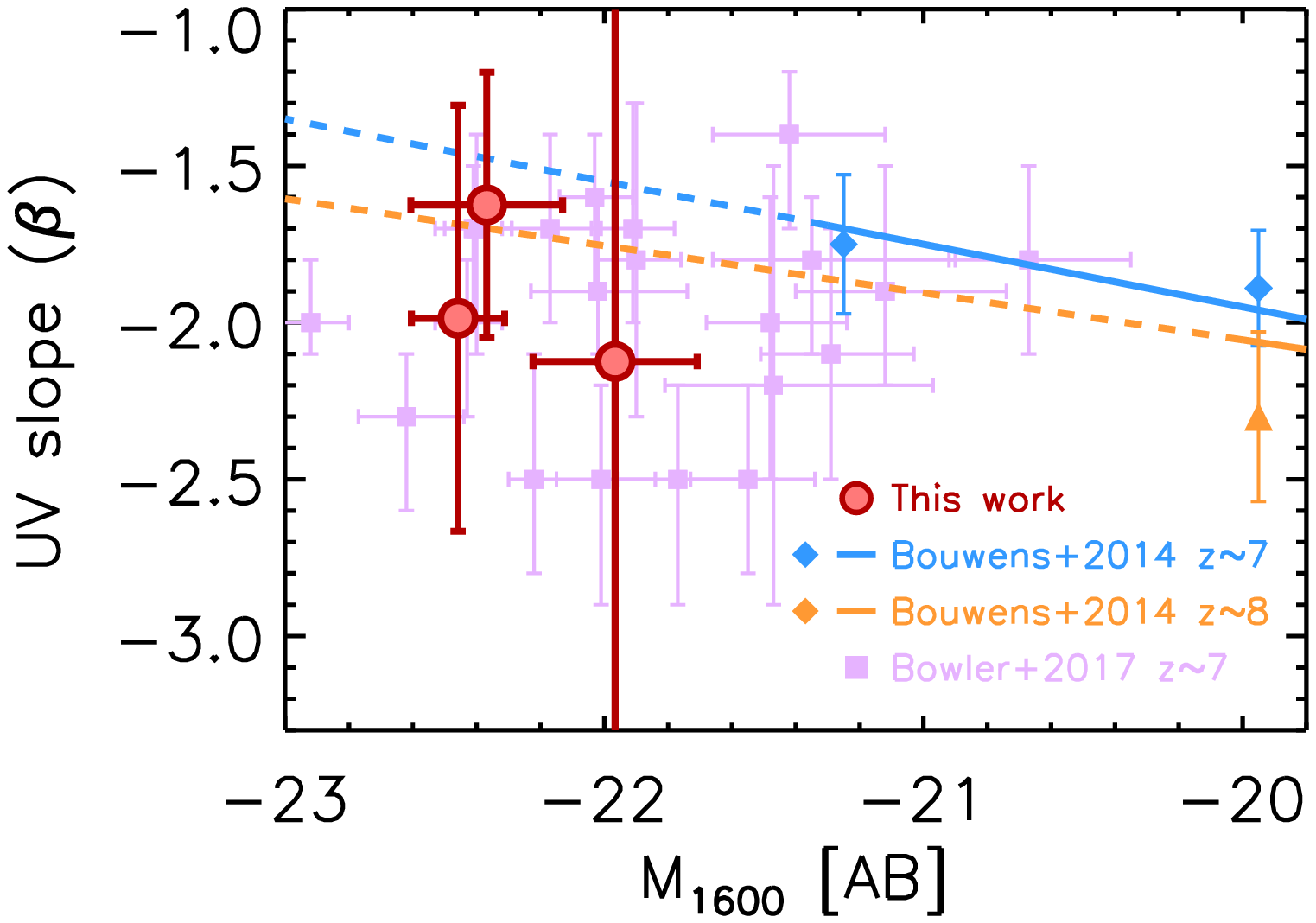}
\caption{Rest-frame UV continuum slope $(\beta)$ versus UV absolute magnitude for the three $z\gtrsim8$ LBGs (red points with errorbars). Individual measurements for the sample of $z\sim7$ LBGs of \citet{bowler2017} are indicated by the  purple points. The  blue and yellow points mark the bi-weight median measurements of \citet{bouwens2014} for the $z\sim7$ and $z\sim8$ samples, respectively. The corresponding best fit relation is shown by the solid blue and yellow lines. The dashed lines indicate the extrapolations of these relations to the luminosities probed in this work. Our measurements of $\beta$ are consistent with the extrapolations of the $z\sim7-8$ relations. \label{fig:beta_MUV}}
\end{figure}

\begin{deluxetable*}{cccccc}
\tablecaption{Physical parameters for the three galaxies with photometric redshift $z\gtrsim8$\tablenotemark{$\mathsection$} \label{tab:properties}}
\tablehead{\colhead{ID} & \colhead{$z_\mathrm{photo}$\tablenotemark{a}} & \colhead{$p(z>6)$}\tablenotemark{b} & \colhead{$M_\mathrm{UV}$\tablenotemark{c}} & \colhead{$\beta$\tablenotemark{d}}  & \colhead{EW$_{0}(H\beta+\mathrm{[OIII]})$\tablenotemark{e}} \\
& & & \colhead{[mag]} & & \colhead{[\AA]}
}
\startdata
UVISTA-Y-1 & $ 8.02^{+0.41}_{-0.49}$ & $0.88$ & $-22.46\pm0.15$ & $-1.98\pm0.67$& $  1041^{+713}_{-515} $\\
UVISTA-Y-5\tablenotemark{$\dagger$} & $ 8.39^{+0.60}_{-0.60}$ & $0.99$ & $-22.37\pm0.24$ & $-1.62\pm0.42$ & $   887^{+1323}_{-686} $ \\
UVISTA-Y-6 & $ 8.35^{+0.65}_{-0.81}$ & $0.83$ & $-21.97\pm0.26$ & $-2.12\pm1.44$ & $  1291^{+1749}_{-940} $ \\
\enddata
\tablenotetext{\mathsection}{The main properties of three other ultra-luminous ($M_\mathrm{UV} \lesssim -22$) $z\sim 8$ candidates are presented in Table~\ref{tab:obs_extra} from Appendix~\ref{app:extra}.}
\tablenotetext{a}{Best photometric redshift  estimate from EAzY, excluding the IRAC bands from the fit,  and corresponding 68\% confidence interval.}
\tablenotetext{b}{Probability, computed by EAzY, that the solution is at $z>6$.}
\tablenotetext{c}{Absolute magnitudes at rest-frame 1600\AA~from EAzY.}
\tablenotetext{d}{Rest-frame UV continuum slopes from the $H_\mathrm{160}$ and UltraVISTA $H$ and $K_\mathrm{S}$ bands.}
\tablenotetext{e}{Rest-frame equivalent width of H$\beta+\mathrm{[OIII]}$ obtained from the $K_\mathrm{S}-[4.5]$ color assuming an SED flat in $f_\nu$ (i.e. $\beta=-2$). If $\beta=-1.8$, the EW and associated uncertainties would be a factor $\sim1.2$ smaller.}
\tablenotetext{\dagger}{Given tension in the flux measurement between the $H_{160}$ and UltraVISTA $H$ bands, we caution the reader that the redshift estimate for this object (and therefore $M_\mathrm{UV}, \beta$ and EW$_0$) is less robust than that of the other two sources in this table. See Sect. \ref{sect:SEDs} and Appendix \ref{app:flux} for further details.}
\end{deluxetable*}

\subsection{$\beta-M_\mathrm{UV}$ relation}
\label{sect:beta}
We measured the rest-frame UV slope ($\beta$) fitting a power-law of the form $f_\lambda \propto \lambda^\beta$ to the fluxes in the $H_\mathrm{160}$ and in the UltraVISTA $H$ and $K_\mathrm{S}$ bands. The results are presented in Figure \ref{fig:beta_MUV} and listed in Table \ref{tab:properties}.  These slopes have an average value of $\beta=-1.91\pm0.26$ and are consistent with the recent determination of the UV slopes of \citet{bowler2017} for LBGs with similar luminosity ($M_\mathrm{UV}\sim-22.5$) identified at $z\sim7$ over the COSMOS/UltraVISTA field, suggesting a slow evolution of $\beta$ for luminous galaxies at early cosmic epochs. Our measurements are also consistent with the UV slope $\beta=-2.1\pm0.3$ from stacking of bright ($M_\mathrm{UV}\sim-21$) $z\sim10$ LBGs by \citet{wilkins2016b}. For comparison, in the plot we also show the bi-weight UV slope measurements at $z\sim7$ and $z\sim8$ from \citet{bouwens2014}, using data from the CANDELS GOODS-N, CANDELS GOODS-S and the \textit{HST} HUDF/XDF fields.

Recent works have identified a correlation between the UV luminosity and the slope of the UV continuum and as a function of redshift: redder slopes are observed at fixed redshift for more luminous galaxies and at fixed luminosity for galaxies at later cosmic times (\citealt{wilkins2011, bouwens2012, finkelstein2012, castellano2012, dunlop2013, jiang2013, bouwens2014, duncan2014, rogers2014, duncan2015}; see also \citealt{oesch2013} and \citealt{stefanon2016} for similar relations of $\beta$ and rest-frame optical luminosities). This behaviour has been interpreted as the emergence of older stellar populations, dust and metals in more luminous galaxies. In Figure \ref{fig:beta_MUV} we also plot the recent determination of the $\beta-M_\mathrm{UV}$  relation of \citet{bouwens2014} at $z\sim7$ and $z\sim8$. Our measurements lie below their \emph{extrapolation} to the luminosity range probed in this work, although the large uncertainties associated to our $\beta$ measurements make them consistent at $<1\sigma$ with those relations, thus preventing from further inspecting any differential evolutionary path of $\beta$ with luminosity and redshift.

\subsection{Size measurement}
The availability of high-resolution imaging from our small  \textit{HST}/WFC3  program allowed us to pursue a first study of the size and morphological properties of extremely bright $z\sim8$ galaxies.

\begin{figure}
\hspace{-0.2cm}\includegraphics[width=8.6cm]{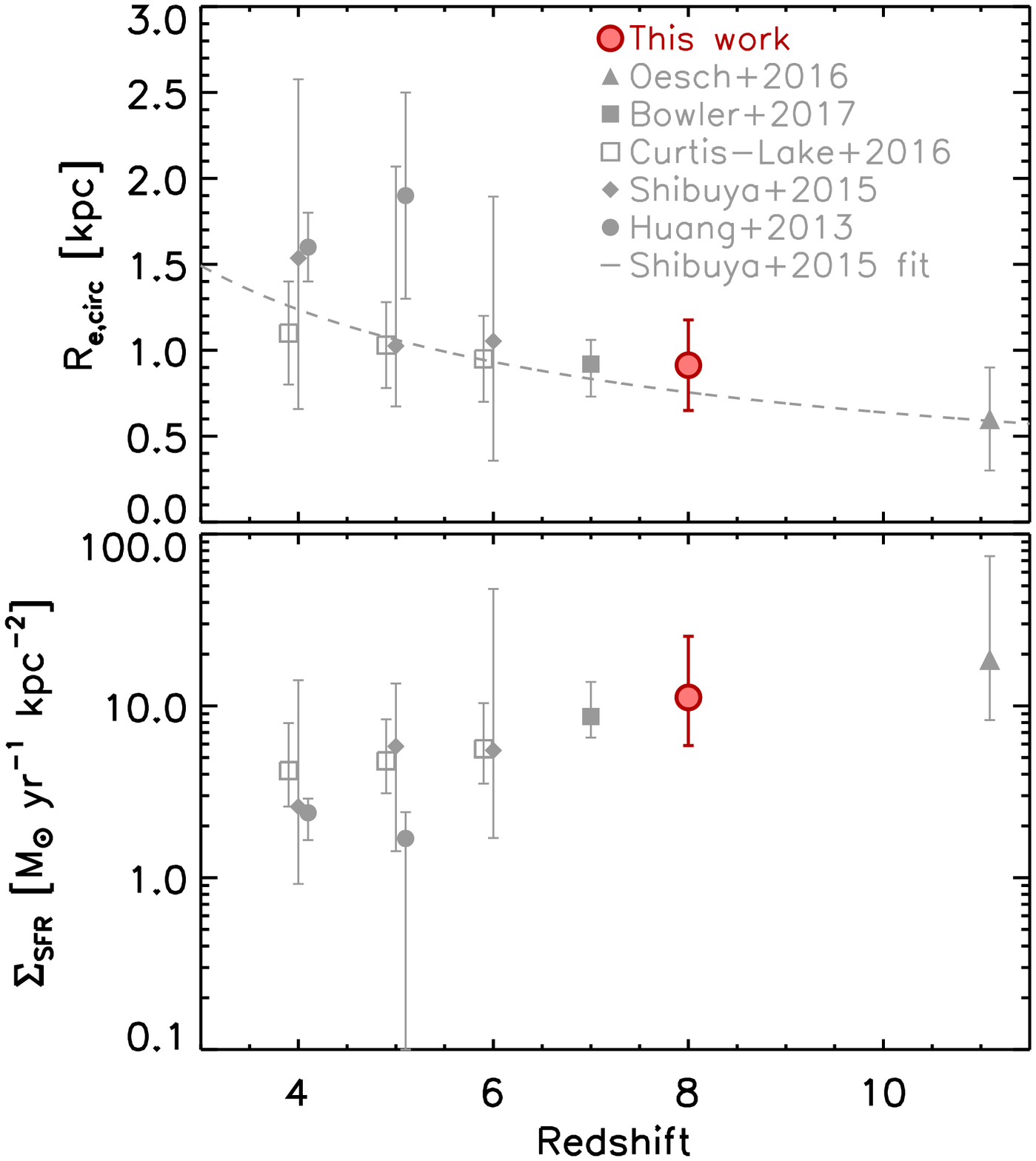}
\caption{{\bf Top panel:} Evolution with redshift of the circularized effective radius of LBGs with $M_\mathrm{UV}\sim-22$~mag. The size we measured for the bright $z\sim8$ galaxy UVISTA-Y-1 is marked by the red filled circle. Also shown are measurement of the $R_{e,\mathrm{circ}}$ for LBGs with $M_\mathrm{UV}\sim-22$ from the literature, as indicated by the legend.  To improve readability, the points of \citet{huang2013} and of \citet{curtis-lake2016} have been arbitrarily shifted by $\Delta z=0.1$ and $\Delta z=-0.1$, respectively. The dashed curve corresponds to the best-fit of \citet{shibuya2015} on the average measurements ($R_{e,\mathrm{circ}} \propto (1+z)^{-0.84}$~kpc).  Given the mild evolution in the characteristic luminosity of the UV LF of LBGs over $z\sim4-10$ and the fact that we are considering the brightest end of the UV LF, the evolution presented here could roughly corresponds to the evolution of the descendants of objects like GN-z11, indicated by the filled triangle at $z\sim11$. {\bf Bottom panel:} Evolution with redshift of the star-formation rate surface density ($\Sigma_\mathrm{SFR}$) for the same sample presented in the top panel, and computed following \citet{ono2013}. The data suggest a (slow) decrease of $\Sigma_\mathrm{SFR}$ with cosmic time for the most luminous galaxies. \label{fig:size_evol}}
\end{figure}

\begin{deluxetable*}{lcccccc}
\tablecaption{Morphological parameters for UVISTA-Y-1 measured on the $JH_{140}$ band with \texttt{galfit}  and \texttt{SExtractor} \label{tab:re}}
\tablehead{\colhead{Algorithm} & \colhead{R.A.} & \colhead{Dec.} & \colhead{$R_{e,\mathrm{circ}}$\tablenotemark{a}}  & \colhead{$q$\tablenotemark{b}} & \colhead{$n$\tablenotemark{c}} & \colhead{$\Sigma_\mathrm{SFR}$\tablenotemark{d}}\\
& \colhead{[J2000]} & \colhead{[J2000]} & [kpc] & & & $[M_\odot$ yr$^{-1}$ kpc$^{-2}$]}
\startdata
\texttt{galfit} & 09:57:47.910 & +02:20:43.50 & $0.9\pm0.3$ & $0.9\pm 0.2$ & 1.5 & $ 11^{+14}_{-5} $ \\
\texttt{SExtractor} &  09:57:47.910 & +02:20:43.50 & $0.7\pm0.1$ & $0.6\pm 0.1$ & $ \cdots$ & $ 17^{+10}_{-5} $ \\
\enddata
\tablenotetext{a}{Circularized effective radius}
\tablenotetext{b}{Minor-to-major axis ratio}
\tablenotetext{c}{S\'ersic index. This was kept fixed when running \texttt{galfit}}
\tablenotetext{d}{Star-formation rate surface density, computed following \citet{ono2013}}
\end{deluxetable*}

Morphological information was recovered running \texttt{galfit} \citep{peng2002, peng2010}, which fits the convolution of a brightness profile with a PSF. The advantage of this approach is that the extracted morphological parameters are  \textit{deconvolved} from PSF effects. For this work, we considered only the \citet{sersic1968} profile, characterized by an effective radius $R_e$ and an index ($n$) expressing how steeply the wings of the profile decrease with the radius. We note here that the symmetry of the brightness profile assumed by the S\`ersic form could result in an over-simplification, and, consequently, limitations, at the time of describing the morphological properties of high redshift galaxies (e.g., $R_e$ in presence of clumpy or merging systems). Indeed, recent studies have shown that sources are observed to be non-symmetric over a wide range of redshifts (e.g., \citealt{law2007, mortlock2013, huertas-company2015, ribeiro2016, bowler2017}), suggesting that high redshift galaxies could be characterized by a range of sizes and morphologies, resulting from different physical processes. 

Considering that the limited S/N of our observations does not allow us to perform a more comprehensive and detailed morphological analysis, we base our analysis on the working hypothesis of a symmetric S\`ersic profile. Furthermore, because of the relatively low signal-to-noise in most of the WFC3 images, for our analysis we only considered the $JH_\mathrm{140}$ band of UVISTA-Y-1 ($\sim$ rest-frame 1600~\AA), i.e., the highest signal-to-noise observation for the brightest object.

The first estimate of the target position, its magnitude, its $R_{e}$, the axis ratio and the value of the local background, needed as input by \texttt{galfit}, was obtained from \texttt{SExtractor}. During the fitting process, we left the $R_{e}$, magnitude and axis ratio free to vary, while we kept the background fixed. Because of the small extension of the brightness profile and of the relatively low signal to noise of our data, during the fitting process we fixed the S\`ersic index to $n=1.5$, consistent with measurements at $z\sim7-10$ (e.g., \citealt{oesch2010, holwerda2015} and \citealt{bowler2017}). We then verified that $R_{e}$ does not systematically change ($\le 10\%$) when the  S\`ersic index varies in the range $1.2<n<2.0$. This variation was added in quadrature to the uncertainty on $R_e$ provided by \texttt{galfit}. In order to ensure the most robust result,  in the fit we also included all the neighbours within $5\farcs0$ from the nominal position of UVISTA-Y-1. Because the $R_e$ directly provided by \texttt{galfit} corresponds to the major semi-axis, and in order to compare to estimates from the literature, we \textit{circularized} it as $R_{e,\mathrm{circ}}=R_e \sqrt{b/a}$, where $b/a$ is the minor-to-major axis ratio. As a consistency check, we also derive $R_e$ using \texttt{SExtractor}. In this case, the final value for $R_{e,\mathrm{SE}}=\sqrt{R_{e,\mathrm{obs, SE}}^2 - r_\mathrm{PSF}^2}$, with $R_{e,\mathrm{obs, SE}}$ the effective radius measured by \texttt{SExtractor} and $R_\mathrm{PSF}$ that of the $JH_{140}$ PSF, with $R_\mathrm{PSF}=0\farcs12$.

We find $R_{e,\mathrm{circ}}=0.9\pm0.2$~kpc from \texttt{galfit}, consistent with $R_{e,\mathrm{SE}}=0.7$ kpc estimated with \texttt{SExtractor}. In Table \ref{tab:re} we list the main morphological parameter we obtain from the two methods. Our values are consistent at $1\sigma$ with estimates of $R_e$ for $M_\mathrm{UV}\sim -22$ LBGs  at $z\sim7$ ($R_{e,\mathrm{circ}}=0.6-0.9\pm0.2$ from a stacking analysis  - \citealt{bowler2017}) and $z\sim11$ ($R_{e,\mathrm{circ}}=0.6\pm0.3$ for the brightest known galaxy at the highest redshift, with luminosity similar to that of our sample - \citealt{oesch2016}). Moreover, because the evolution of the characteristic luminosity of the UV LF is small for $z\sim4-10$  (\citealt{bouwens2015, finkelstein2015a, bowler2017, ono2017}), and considering that our sources constitute the very bright end of the UV LF, the absolute magnitude corresponding to a constant cumulative number density should evolve very little over $z\sim4-10$. This means that, under the further assumption of a smooth evolution of the star-formation history (SFH), selecting galaxies with approximately the same (high) luminosity corresponds to selecting the descendants of the luminous galaxies observed at the highest redshift in the sample. 

In the top panel of Figure \ref{fig:size_evol}  we present a compilation of size measurements  for LBGs at $z>4$ and $M_\mathrm{UV}\sim-22$ from \citet{huang2013, shibuya2015, oesch2016, curtis-lake2016} and \citet{bowler2017}. The plot suggests only a modest evolution  in size for luminous galaxies (factor of $\sim3\times$) during approximately the first 1.5 Gyr of cosmic time. The bottom panel of Figure \ref{fig:size_evol} presents the evolution of the star-formation rate (SFR) surface density ($\Sigma_\mathrm{SFR}$), computed using the recipe of \citet{ono2013}. The SFR is estimated from the UV luminosity following \citet{kennicutt1998} under the assumption of negligible dust obscuration. The SFR  is then divided by the area corresponding to $R_{e,\mathrm{circ}}$ and applying a further factor $0.5$ to take into account that observationally we can only access approximately half of the surface of each galaxy. The value we find for $\Sigma_\mathrm{SFR}\sim11 M_\odot$ yr$^{-1}$ kpc$^{-2}$ is consistent with measurements at lower luminosities (e.g., \citealt{ono2013, holwerda2015,shibuya2015}, although \citet{oesch2010} found $\Sigma_\mathrm{SFR}$ for $L\lesssim L^*$ galaxies a factor $\sim3\times$ lower). 

Interestingly, but unsurprisingly, $\Sigma_\mathrm{SFR}$ decreases with cosmic time, although with marginal statistical significance. Some recent studies of $z\sim4-8$ LBGs have found indication for a non-evolving $\Sigma_\mathrm{SFR} - z$ relation (e.g., \citealt{oesch2010, ono2013}). This is qualitatively consistent with the increase of the star-formation rate density with cosmic time combined with the increase in size.  Our (mildly) evolving $\Sigma_\mathrm{SFR}$, instead, is the direct consequence of the evolution in size of galaxies with luminosity approximately constant over $4\lesssim z \lesssim 10$. 

We finally note that recent methods for the morphological analysis of high-redshift galaxies have found that the evolution of size could have been much less pronounced than recovered through more classical approaches (e.g., \citealt{law2007, curtis-lake2016, ribeiro2016}). While there is no reason to exclude this could be the case at even higher redshift, data with better S/N is necessary for a more robust assessment.

\subsection{Volume density at $z\sim8$}
\label{sect:LF}

\begin{figure}
\includegraphics[width=8cm]{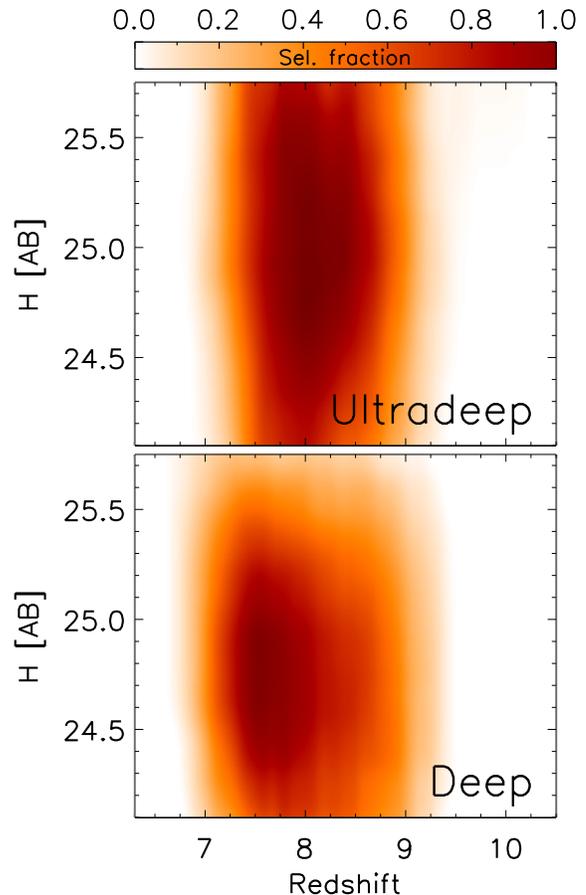}
\caption{Selection functions corresponding to the criteria adopted to identify candidate $z\sim8$ LBGs, computed with a Monte Carlo simulation on real data. The top panel refers to the ultradeep stripes, while the lower panel to the UltraVISTA deep stripes. The original un-evenly spaced measurements have been smoothed with a boxcar filter of 5 pixels, after casting them onto a regular grid for display purposes. Darker regions correspond to higher selection rates as indicated by the scale at the top of the figure. Our criteria allow us to select galaxies at $7.1 \lesssim z \lesssim 9.2$ and $6.9 \lesssim z \lesssim 9.3$, respectively for the ultradeep and deep regions. \label{fig:sel_function}}
\end{figure}

Using the results obtained in the previous sections, we estimate the contribution of the three candidate $z\sim8$ LBGs to the UV LF.  Here we focus on the HST sample analyzed in this work, which constitute the brightest end of the UV LF. A more comprehensive UV LF including the complete sample of fainter sources detected over COSMOS/UltraVISTA will be presented in \citet{labbe2017}.   The measurement of the volume density relies on estimating the detection completeness and the selection function associated to our selection criteria.  We recovered these two quantities using similar procedures as described in \citet{bouwens2015}. Briefly, we generated catalogs of mock sources with realistic sizes and morphologies by randomly selecting images of $z\sim4$ galaxies from the Hubble Ultra Deep Field (\citealt{beckwith2006, illingworth2013}) as templates. The images were re-sized to account for the change in angular diameter distance with redshift and for evolution of galaxy sizes at fixed luminosity  (effective radius $r_e\propto (1+z)^{-1}$: \citealt{oesch2013, ono2013, holwerda2015, shibuya2015}). The template images were then inserted into the observed images, assigning colors expected for star forming galaxies in the range $6<z<11$. The colors were based on a UV continuum slope distribution of $\beta=-1.8 \pm 0.3$ to match the measurements for luminous $6<z<8$ galaxies and consistent with the determinations from this work (\citealt{bouwens2012, finkelstein2012, rogers2014}). The simulations include the full suite of \textit{HST}, ground-based, and \textit{Spitzer}/IRAC images. For the ground-based and Spitzer/IRAC data the mock sources were convolved with appropriate kernels to match the lower resolution PSF. To simulate IRAC colors we assume a continuum flat in $f_\nu$ and emission lines with fixed rest-frame EW($H\alpha$+[NII]+[SII]) = 300\AA~and rest-frame EW([OIII]+$H\beta$) = 500\AA, consistent with the results of \citet{labbe2013, stark2013, smit2014, smit2015, rasappu2016}. The same detection and selection criteria as described in Sect. \ref{sect:selection} were then applied to the simulated images to recover the completeness as a function of magnitude and the selection as a function of magnitude and redshift.

Given that the source detection was performed on the UltraVISTA mosaics, roughly characterized by a dual depth (ultradeep and deep), the above process was independently executed in regions corresponding to the two depths. Figure \ref{fig:sel_function} presents the selection functions associated to our criteria for the UltraVISTA ultra-deep and deep stripes, used to estimate the co-moving volumes entering the LF determinations. The plots show that in the ultradeep stripes our criteria allow us to select galaxies at $7.1\lesssim z \lesssim 9.2$. In the deep stripes, instead, the range of redshift selection is slightly broader, $6.9\lesssim z \lesssim 9.3$, qualitatively consistent with the fact that shallower depths in the NIR bands can also accommodate slightly different solutions.

\begin{figure}
\hspace{-0.4cm}\includegraphics[width=9.5cm]{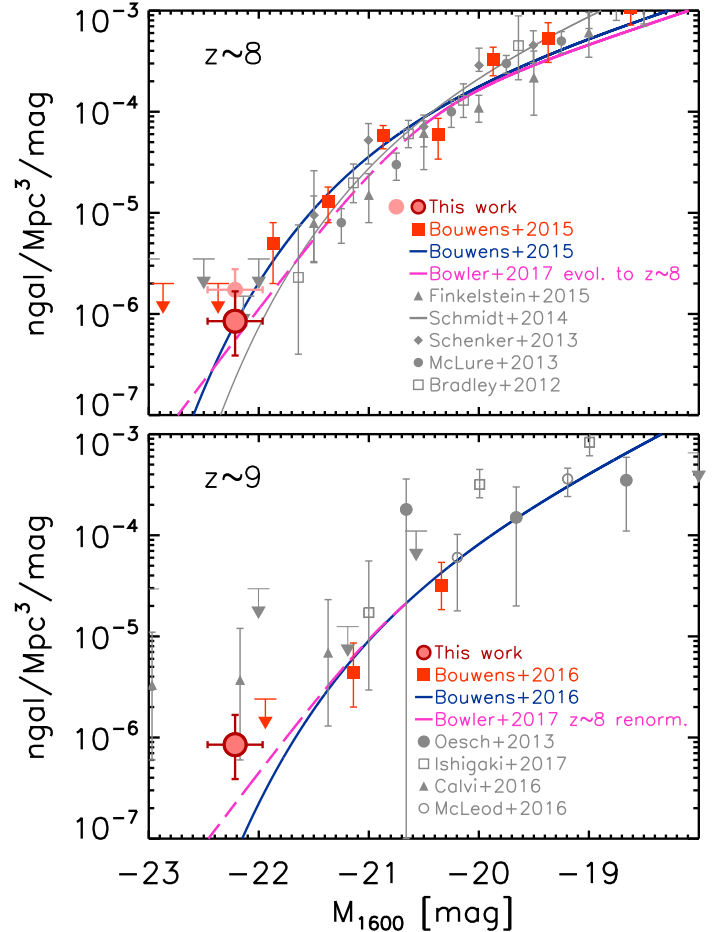}
\caption{{\bf Top panel:} The red point with errorbars marks our estimate of volume density associated to the sample of candidate luminous $z\gtrsim8$ galaxies considered in this work. The pink filled circle corresponds to the volume density after adding to the sample three luminous $z\sim8$ LBGs from \citet{labbe2017}, not targeted in the \textit{HST} proposal. Recent UV LF determinations at $z\sim8$ from  \citet{bradley2012,mclure2013,schenker2013, schmidt2014, bouwens2015,finkelstein2015a} are also reported, with plotting conventions detailed by the legend. The blue curve corresponds to the Schechter form of \citet{bouwens2015}, while the magenta curve represents the evolution to $z\sim8$ of the $z\sim7$ DPL of \citet{bowler2017} according to \citet{bouwens2016}. {\bf Bottom panel:} Here we compare our volume density estimate (red point) to measurements of the UV LF at $z\sim9$ \citep{oesch2013, bouwens2016, calvi2016, mcleod2016, ishigaki2017}. The blue curve represents the Schechter function from \citet{bouwens2016}. The magenta curve presents the bright-end of the dual power law from \citet{bowler2017} evolved to $z\sim8$ following \citet{bouwens2016} whose characteristic density has been adjusted to match that of the Schechter function at the characteristic luminosity.  \label{fig:LF}}
\end{figure}

The volume density associated to the three $z\gtrsim8$ candidate LBGs was computed using the $1/V_\mathrm{max}$ method (\citealt{schmidt1968}), and following the prescription of \citet{avni1980} for a \emph{coherent} analysis, in order to deal with the different depths of the deep and ultradeep stripes of the UltraVISTA field. The $1/V_\mathrm{max}$ method is intrinsically sensitive to local overdensities of galaxies; however, given the small sample considered in this work, we consider its potential effects by including the cosmic variance in the error budget. On the other hand, the $1/V_\mathrm{max}$ method directly provides the normalisation of the LF. 

Considering that the absolute magnitudes of the three $z\gtrsim8$ candidate LBGs are within 0.5~mag, the volume density was computed in one bin only. We obtain a volume density of $\Phi=8.49^{+8.23}_{-4.60}\times 10^{-7}$~Mpc$^{-3}$~mag$^{-1}$ at $M_\mathrm{UV}=-22.21\pm 0.25$. The uncertainties associated to the volume density were computed following the recipe of \citet{gehrels1986}, and adding in quadrature 24\% of cosmic variance following \citet{moster2011}. Our measurement is shown in the top panel of Figure \ref{fig:LF} with a filled red circle, together with a compilation of previous determinations of the bright end of the UV LF at $z\sim8$. To avoid potential systematics, we limit our comparison to studies based on field galaxies, excluding UV LFs from samples based on  galaxy cluster (\citealt{bradley2012,mclure2013,schenker2013,schmidt2014,bouwens2015,finkelstein2015a}). Our measurement constitutes the first volume density estimate for $M_\mathrm{UV}\lesssim-22$ at $z\sim8$ with confidence $\gtrsim 1\sigma$ and is consistent with previous upper limits. In the same panel we also reproduce the \citet{schechter1976} parameterization of the UV LF at $z\sim 8$ from \citet{bouwens2015}. Our estimate of the bright end agrees well with the exponential decline of the current Schechter form. Since the sample of \citet{labbe2017} includes three more potential galaxies at $7.5\lesssim z_\mathrm{phot}\lesssim8.5$, which however did not enter the selection criteria for the \textit{HST} proposal, here we also present the volume density obtained including all the six sources. The multi-wavelength photometry and results of SED fitting for these three additional sources  are presented in Appendix \ref{app:extra}. At $M_\mathrm{UV}=-22.21\pm 0.25$ the volume density is $\Phi=17.3^{+10.3}_{-6.9}\times 10^{-7}$~Mpc$^{-3}$~mag$^{-1}$. This measurement is plotted in Figure \ref{fig:LF} with a pink filled circle and it is still consistent with the recent determinations of the bright end of the $z\sim8$ UV LF (s).

Recent studies of the bright end of the UV LF at $z\gtrsim6$ suggest that the LF could be parameterised by a double power law (DPL - \citealt{bowler2014,bowler2015,bowler2017,ono2017}) originated by an excess of luminous galaxies compared to the exponential decline of the Schechter function. The magenta curve in Figure \ref{fig:LF} presents the DPL of \citet{bowler2017} after evolving the faint-end slope, characteristic magnitude and normalization factor to $z\sim8$ using the evolution of \citet[see their Sect. 5.1]{bouwens2015}. The DPL well describes the points, in particular considering the measurements of \citet{mclure2013} at absolute magnitudes brighter than $L^*$. Our single measurement is not able to distinguish between the two scenarios, though, as the corresponding absolute magnitude lies at the intersection of the Schechter and the DPL forms.

Because the $z_\mathrm{phot}$ solutions for half of our sample of candidate $z\gtrsim8$ LBGs may have values close to $z_\mathrm{phot}\sim9$ when including the IRAC bands, we also computed the volume density associated to the three sources (UVISTA-Y-5, UVISTA-Y-6 and UVISTA-Y-2) with $z_\mathrm{phot}>8.5$. We obtain $\Phi=8.5^{+8.2}_{-4.6}\times 10^{-7}$~Mpc$^{-3}$~mag$^{-1}$.  In the lower panel of Figure \ref{fig:LF} we compare our $z\sim9$ volume density measurement with the UV LFs at $z\sim9$ from \citet{oesch2013, mclure2013, bouwens2016, calvi2016, mcleod2016, ishigaki2017}. Our estimate is consistent with the measurement of \citet{calvi2016}, although it corresponds to higher densities than expected from the Schechter determination of \citet{bouwens2016}. In the same panel we also plot the bright-end of the dual power-law we constructed for the $z\sim8$ bin, renormalized to match the density of the Schechter form at the characteristic magnitude. It agrees within the error bars with our volume measurement.

The volume density we estimate at $z\sim8$ is consistent with that at $z\sim9$, albeit the large statistical uncertainties, and suggests a slow evolution of the brightest objects at early cosmic epochs. Remarkably, this is still valid  considering that our volume density measurements are consistent at $\sim 1\sigma$ with the volume density estimate for the $z\sim11$ source GN-z11 (\citealt{oesch2016}). Assuming a smooth SFH, this could imply that these bright (and possibly massive) galaxies assembled extremely rapidly in the first few hundreds Myr after the Big Bang. A very bursty SFH, instead, would make any interpretation challenging, because the number density would be a (random) combination of bright (massive) galaxies with reduced SFR and lower mass galaxies with strong SFR.

\section{Summary and Conclusions}

Here we report on \textit{HST} WFC3/IR observations on five very bright $z\sim8-9$ candidates identified over UltraVISTA. The targeted sources were drawn from a sample of 16 very bright $z\sim8-9$ galaxies identified by \citet{labbe2017}, and constituted the brightest sources from that sample ($24.5<H<25.2$) with a plausible $z\gtrsim8.5$ solution.  The five sources in this sample (labelled UVISTA-Y-1, UVISTA-Y-5, UVISTA-Y-6, UVISTA-J-1 and UVISTA-J-2) that stood out for their brightness ($24.5\lesssim H\lesssim 25.2$), for having plausible redshift $z_\mathrm{phot}\sim 8.5$ solutions, for being positioned over the UltraVISTA deep stripes and had coverage from the deepest optical ground-based data  have recently been observed with \textit{HST}/WFC3 (PI: R. Bouwens, PID: 14895) to try to confirm their nature. 

The present work is devoted to the analysis of those sources specifically targeted with \textit{HST}/WFC3 follow-up observations.  Nevertheless, this analysis does present three other ultra-luminous $z\sim8$ galaxies from the \citet{labbe2017} UltraVISTA selection in Appendix~\ref{app:extra} -- since they play a role in our volume density determination -- such that this paper includes the full set of properties for the six most luminous $z\sim8$ sources identified over UltraVISTA.  Full details on the sample assembly and analysis of the complete sample are presented in \citet{labbe2017}.

The \textit{HST}/WFC3 observations were performed in the $z_\mathrm{098}$, $J_\mathrm{125}$ and $H_\mathrm{160}$ bands (Figure \ref{fig:cutouts}) in March 2017, for a total of 1 orbit per source, reaching depths of $\sim 25.8, 25.0, 24.5$~mag, respectively. One source (UVISTA-Y-1) also benefits from the archival data of the program SUSHI (PI: N. Suzuki - PID: 14808), which provides coverage in the $Y_\mathrm{105}$ and $JH_\mathrm{140}$ bands to $25.6$ and $25.5$~mag, respectively ($5\sigma$, $1\farcs2$ aperture diameter - Figure \ref{fig:Y1_cutouts}). Leveraging the new HST images, we reprocessed the existing ground and space-based data, extracting accurate flux measurements with the \texttt{mophongo} software (\citealt{labbe2006, labbe2010a, labbe2010b, labbe2013, labbe2015}). In our analysis we also now included the recently released ground based $g, r, i, z$ and $Y$ data from the ultradeep layer of the HyperSuprimeCam survey (\citealt{aihara2017a, aihara2017b}).

Our analysis confirms the photometric redshift of three sources (UVISTA-Y-1, UVISTA-Y-5 and UVISTA-Y-6) to be $8.0<z_\mathrm{photo}<8.7$ (Figure \ref{fig:z9}). Their measured luminosity $M_\mathrm{UV}\sim -22.3$ makes them perhaps the brightest, most reliable galaxies at $z \sim 8-9$ identified to date. The uniquely deep optical, near-IR, and Spitzer/IRAC data available for these  sources is the reason for our high confidence in
their nature (Figure \ref{fig:zMUV}). However, our analysis also demonstrates that the remaining two sources (UVISTA-J-1 and UVISTA-J-2) are very likely lower redshift interlopers, with nominal redshifts of $z_\mathrm{phot}\sim2$ (Figure  \ref{fig:z2}).

The three $z\sim8$ candidate LBGs are characterized by average UV continuum slopes $\beta=-1.91\pm0.26$, consistent with lower redshift ($z\sim7$), similarly bright samples of LBGs of \citet{bowler2017}, suggesting a differential evolution of $\beta$ for the most luminous galaxies compared to $L^*$ or sub-$L^*$ galaxies at early cosmic epochs. Our $\beta$ are bluer than the extrapolations of measurements for lower luminosity LBGs from CANDELS data, although the large uncertainties make them consistent at $<1\sigma$ (e.g., \citealt{bouwens2014} - Figure \ref{fig:beta_MUV}), preventing from deriving any further conclusion on differential evolution. 

For our bright source UVISTA-Y-1 we measure a size of $R_{e,\mathrm{circ}}=0.9\pm0.2$~kpc, consistent with sizes of similarly luminous LBGs at $z\sim7-10$, and suggesting very mild evolution over the first 1.5~Gyr of cosmic time (Figure \ref{fig:size_evol}).

Finally, using the $1/V_\mathrm{max}$ formalism of \citet{avni1980}, we computed the volume density $\Phi$ associated to the three candidate $z\sim8$ LBGs. We find $\Phi=8.49^{+8.23}_{-4.60}\times 10^{-7}$~Mpc$^{-3}$~mag$^{-1}$ at $M_\mathrm{UV}=-22.21\pm 0.25$. This constitutes the first measurement of the number density of $M_\mathrm{UV} < -22$~mag LBGs at $z\sim8$ based on actual detection of sources. We also estimate the volume density of star-forming galaxies at $z\sim8$, including the full sample of six $M_\mathrm{UV} \lesssim -22$~mag galaxies identified in \citet{labbe2017} -- three of which are presented in Appendix \ref{app:extra}. Together with the three candidate $z\sim8$ LBGs confirmed by our main analysis, they constitute a complete sample of $z\sim8$ LBGs with $M_\mathrm{UV} \lesssim -22$~mag. The measured volume density associated to this complete sample is  $\Phi=17.3^{+10.3}_{-6.9}\times 10^{-7}$~Mpc$^{-3}$~mag$^{-1}$ at $M_\mathrm{UV}=-22.21\pm 0.25$ mag.

 Unfortunately, given the large statistical uncertainties, we cannot use our current constraints on the bright end of the LF to discriminate between a Schechter or double power-law form. Improvements on this front could come either from the detection of sources at even higher luminosities, where the discrepancies between the Schechter and the double power-law form are larger, or from increased samples of galaxies within the current luminosity ranges,  reducing the statistical uncertainties on the LF measurements. \\

This work further stresses the importance of the high-resolution imaging provided by \textit{HST} in the study of the galaxy populations at early cosmic epochs, enabling to refine the photometric redshifts and indentify interlopers, resulting in cleaner samples. Our results, however, are based on photometric redshifts from broad-band imaging. A more robust picture inevitably requires spectroscopic confirmation. To this aim, we have started spectroscopic follow-up with Keck/MOSFIRE and VLT/X-Shooter. The sensitivities at observed optical/NIR wavelengths make it challenging with the current instrumentation. Bright high-redshift objects like those analysed in this work offer two possibilities: 1) they constitute prime candidates for \textit{future} spectroscopic follow-up with JWST; 2) their brightness together with refined photometric redshifts suggests them as valid targets for \textit{current} ALMA observations, possibly resulting in spectroscopic confirmation even before the start of operations of JWST.

\acknowledgements \textit{Acknowledgements:} MS would like to thank Pieter van Dokkum, Jorryt Matthee and Corentin Schreiber for useful discussion. KIC acknowledges funding from the European Research Council through the award of the Consolidator Grant ID 681627-BUILDUP. Based on data products from observations made with ESO Telescopes at the La Silla Paranal Observatory under ESO programme ID 179.A-2005 and on data products produced by TERAPIX and the Cambridge Astronomy Survey Unit on behalf of the UltraVISTA consortium. Based in part on data collected at the Subaru Telescope and retrieved from the HSC data archive system, which is operated by Subaru Telescope and Astronomy Data Center at National Astronomical Observatory of Japan. The Hyper Suprime-Cam (HSC) collaboration includes the astronomical communities of Japan and Taiwan, and Princeton University. The HSC instrumentation and software were developed by the National Astronomical Observatory of Japan (NAOJ), the Kavli Institute for the Physics and Mathematics of the Universe (Kavli IPMU), the University of Tokyo, the High Energy Accelerator Research Organization (KEK), the Academia Sinica Institute for Astronomy and Astrophysics in Taiwan (ASIAA), and Princeton University. Funding was contributed by the FIRST program from Japanese Cabinet Office, the Ministry of Education, Culture, Sports, Science and Technology (MEXT), the Japan Society for the Promotion of Science (JSPS), Japan Science and Technology Agency (JST), the Toray Science Foundation, NAOJ, Kavli IPMU, KEK, ASIAA, and Princeton University.  This paper makes use of software developed for the Large Synoptic Survey Telescope. We thank the LSST Project for making their code available as free software at  http://dm.lsst.org. The Pan-STARRS1 Surveys (PS1) have been made possible through contributions of the Institute for Astronomy, the University of Hawaii, the Pan-STARRS Project Office, the Max-Planck Society and its participating institutes, the Max Planck Institute for Astronomy, Heidelberg and the Max Planck Institute for Extraterrestrial Physics, Garching, The Johns Hopkins University, Durham University, the University of Edinburgh, QueenÕs University Belfast, the Harvard-Smithsonian Center for Astrophysics, the Las Cumbres Observatory Global Telescope Network Incorporated, the National Central University of Taiwan, the Space Telescope Science Institute, the National Aeronautics and Space Administration under Grant No. NNX08AR22G issued through the Planetary Science Division of the NASA Science Mission Directorate, the National Science Foundation under Grant No. AST-1238877, the University of Maryland, and Eotvos Lorand University (ELTE) and the Los Alamos National Laboratory. Based on observations obtained with MegaPrime/MegaCam, a joint project of CFHT and CEA/IRFU, at the Canada-France-Hawaii Telescope (CFHT) which is operated by the National Research Council (NRC) of Canada, the Institut National des Science de l'Univers of the Centre National de la Recherche Scientifique (CNRS) of France, and the University of Hawaii. This work is based in part on data products produced at Terapix available at the Canadian Astronomy Data Centre as part of the Canada-France-Hawaii Telescope Legacy Survey, a collaborative project of NRC and CNRS. This research has made use of the NASA/ IPAC Infrared Science Archive, which is operated by the Jet Propulsion Laboratory, California Institute of Technology, under contract with the National Aeronautics and Space Administration.

\appendix
\section{Inconsistent flux measurements and interlopers}
\label{app:flux}

Our analysis contains three sources (UVISTA-Y-5, UVISTA-J-1 and UVISTA-J-2) which have flux measurements in the $H_{160}$ and UltraVISTA $H$ bands that are inconsistent at $\gtrsim 2 \sigma$. For UVISTA-J-1 and UVISTA-J-2, the lower flux measurement in the $H_{160}$  contributes to favouring a low-redshift solution. In this section we discuss possible reasons for this systematic offset. Specifically we consider photometric zero-point offsets, time variability, nebular emission and extended morphology.

When beginning the analysis for this paper, we had checked the zero point of the UltraVISTA DR3 $H$-band mosaic comparing the total flux in the $H_{160}$ and UltraVISTA $H$ bands of  $\sim50$ bright unsaturated point sources, identified over the footprint of the CANDELS/COSMOS field. This region is covered by deep WFC3 data from the CANDELS program and by one of the ultradeep stripes of the UltraVISTA DR3 release, ensuring the highest S/N in both bands. The total flux was recovered from the curve-of-growth, measured out to $10\farcs0$ radii. In doing so, we masked any detected source, other than the point source itself, inside the radius adopted for the measurement. On this basis, we revised the UltraVISTA ZP estimates faintward by 0.1 mag.   This zeropoint offset is already included in the photometry we provide in Table \ref{tab:photometry}.

In principle, the different releases of the UltraVISTA dataset allow us to explore potential variations of source flux with time.  We find that the photometry of UVISTA-J-1 in the UltraVISTA $H$ band at the two epochs corresponding to DR1 and DR3 is consistent within uncertainties. Unfortunately, for UVISTA-Y-5 and UVISTA-J-2 this check is unfeasible, because the sources are undetected in the DR1 mosaic. While for UVISTA-Y-5 a non-detection in the DR1 data is still consistent at $\sim 2\sigma$ with the flux measured on the DR3 mosaic, this is not the case for UVISTA-J-2, whose flux should be $\sim1.7\times$ higher than UVISTA-Y-5, hence suggesting potential variability.
 
The coverage of the UltraVISTA $H$-band filter extends, at redder wavelengths, $\sim 10^3$ \AA~beyond that of the \textit{HST}/WFC3 $H_{160}$. Red $H_{160}-H$ colors could then reflect the presence of nebular emission whose observed wavelength falls in the extra $\sim 10^3$  \AA~covered by the UltraVISTA $H$ band. At $z\sim8$ that wavelength range contains CIII]$\lambda\lambda 1907, 1909$ \AA. The resulting equivalent width would be in excess of $\sim 500$ \AA, physically unlikely (see e.g., \citealt{stark2017} who found EW$_0\sim 22$ \AA~for CIII] at $z\sim 7$). Even assuming that CIII] is dominating the flux in the UltraVISTA $H$ band,  similar EW for CIV  would be required to explain the flux in the $K_\mathrm{S}$ band, together with a very red SED red-ward of $K_\mathrm{S}$ to match the fluxes in the $3.6\mu$m and $4.5\mu$m bands.  One further possibility could be that the observed discrepancy was originated by [OIII]+H$\beta$ nebular lines if the galaxy had a redshift $z\sim 2.4-2.6$. In this case, EW$_0$([OIII]+H$\beta$)$\gtrsim 1300$ \AA. This value is larger than inferred from conversions of observed H$\alpha$ EW at $z\sim2$ (e.g., \citealt{erb2006}); this solution becomes even more unlikely considering that the best-fit template required to match $z\sim2$ is characterized by old age and dust attenuation. The old age would be inconsistent with strong nebular emission; furthermore the dust attenuation would make the unreddened EW even larger. We conclude that nebular emission is not the likely origin of the observed discrepancy.

If  the sources were characterized by extended, low surface-brightness wings, the short exposures in the WFC3 bands would not be enough to detect them, hence reducing their estimated luminosity. Unfortunately, a stacking of the WFC3 bands did not present any evidence for extended wings, possibly because of the low S/N characterizing our \textit{HST} observations.\\
 
Visual inspection of the UltraVISTA mosaics showed that  UVISTA-J-2 has a very bright ($\approx 12$~mag ) neighbouring star located $\sim40$ arcsec north-east. Even though the procedure we adopted to extract the flux measurements takes care of estimating the background, we can not exclude a residual contamination from the wings of the bright source. It is remarkable, though, that if we exclude the UltraVISTA data from the SED of UVISTA-J-2, the \textit{HST} photometry is consistent with the SED of an LBG at $z\sim9.5$.  Similarly, if we exclude the UltraVISTA data from UVISTA-J-1, we obtain an SED consistent with an LBG at $z\sim9.5$, although less robust than UVISTA-J-2 given the detection in the HSC $z$ band. In conclusion this suggests that for these two objects a high-redshift solution is not completely ruled out.

As we show in Figure \ref{fig:z2}, the WFC3 observations are responsible for or contribute to the solution at $z\sim2$.  The above considerations stress the importance of performing high S/N \textit{HST} follow-up of the candidate bright LBGs detected from ground-based surveys, in order to reduce the systematic uncertainties in the photometry and produce more stable photometric redshift measurements. The current samples of $z\ge4$ LBGs at lower luminosities are generally built from deep \textit{HST} imaging. The higher S/N of the \textit{HST} observations in these fields greatly reduces the chances of uncertain redshifts identification of their redshifts. Nonetheless, issues in the assessment of the nature of candidate LBGs arise at the faint end of the UV LF (see e.g. \citealt{bouwens2017}).
 
 \begin{figure*}
\includegraphics[width=18cm]{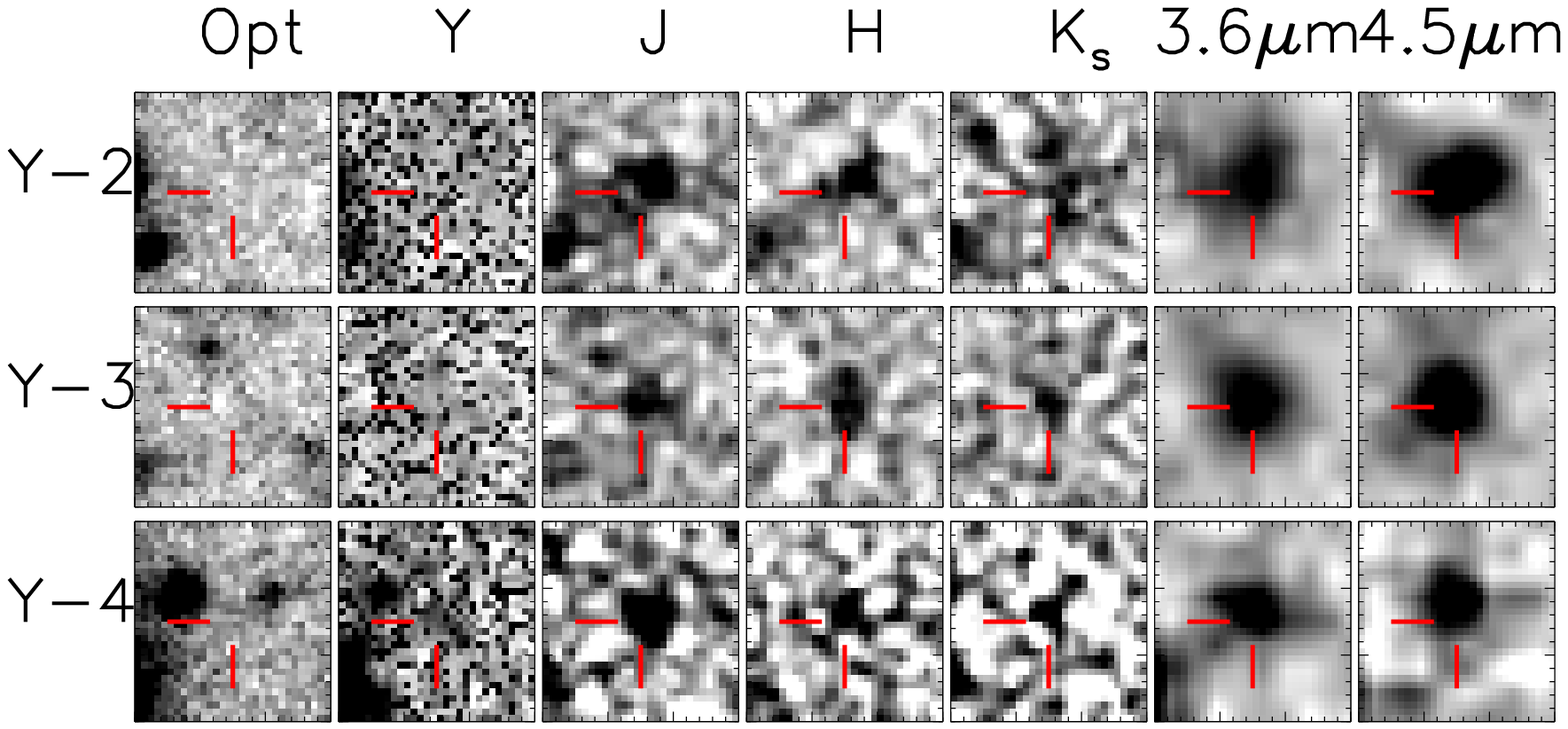}
\caption{Image stamps in inverted grey scale of the three additional bright candidate $z\gtrsim8$ LBGs in the stacked optical, stacked $Y$, UltraVISTA $J, H$ and $K_\mathrm{S}$ and \textit{Spitzer} IRAC $3.6\mu$m and $4.5\mu$m bands. Each row corresponds to a source, as labeled on the left, where we omitted the prefix UVISTA- from the object name for clarity. Each cutout is $5\farcs0  \times 5\farcs0$. \label{fig:cuts_extra}}
\end{figure*}

 \begin{deluxetable*}{lr@{ $\pm$}rr@{ $\pm$}rr@{ $\pm$}r}
\tablecaption{Total flux densities for the three additional candidate $z\gtrsim8$ LBGs over COSMOS/UltraVISTA included in the estimate of the LF. \label{tab:phot_extra}}
\tablehead{\colhead{Filter} & \twocolhead{UVISTA-Y-2} & \twocolhead{UVISTA-Y-3} & \twocolhead{UVISTA-Y-4}  \\
& \twocolhead{[nJy]} & \twocolhead{[nJy]} & \twocolhead{[nJy]} 
}
\startdata
            CFHTLS $u^*$  & $      -6 $  & $      15 $  & $      -6 $  & $      15 $  & $     -10 $  & $      15 $  \\
                 SSC $B$  & $      -1 $  & $      10 $  & $      -8 $  & $       8 $  & $       5 $  & $       9 $  \\
                 HSC $g$  & $       2 $  & $      17 $  & $     -11 $  & $      17 $  & $      -8 $  & $      17 $  \\
              CFHTLS $g$  & $      -1 $  & $      13 $  & $     -12 $  & $      13 $  & $       2 $  & $      13 $  \\
                 SSC $V$  & $     -14 $  & $      22 $  & $     -14 $  & $      19 $  & $      29 $  & $      20 $  \\
                 HSC $r$  & $      -4 $  & $      17 $  & $       0 $  & $      16 $  & $       7 $  & $      17 $  \\
              CFHTLS $r$  & $     -12 $  & $      19 $  & $      -4 $  & $      19 $  & $       7 $  & $      20 $  \\
               SSC $r^+$  & $     -15 $  & $      20 $  & $      -5 $  & $      17 $  & $      10 $  & $      18 $  \\
              SSC  $i^+$  & $    -139 $  & $      89 $  & $     -77 $  & $     110 $  & $    -103 $  & $      90 $  \\
              CFHTLS $y$  & $     -18 $  & $      25 $  & $       3 $  & $      24 $  & $     -14 $  & $      25 $  \\
               CFHTLS $i$ & $     -28 $  & $      26 $  & $     -30 $  & $      25 $  & $       3 $  & $      26 $  \\
                 HSC $i$  & $     -44 $  & $      24 $  & $     -11 $  & $      24 $  & $     -10 $  & $      24 $  \\
              CFHTLS $z$  & $     -13 $  & $      55 $  & $     -38 $  & $      54 $  & $       7 $  & $      54 $  \\
                 HSC $z$  & $      -1 $  & $      35 $  & $     -36 $  & $      35 $  & $     -50 $  & $      35 $  \\
              SSC  $z^+$  & $    -102 $  & $      75 $  & $     -10 $  & $      73 $  & $      43 $  & $      71 $  \\
                 HSC $y$  & $      31 $  & $      85 $  & $     -89 $  & $      86 $  & $      33 $  & $      84 $  \\
              UVISTA $Y$  & $      29 $  & $      53 $  & $      88 $  & $      53 $  & $     173 $  & $      54 $  \\
             UVISTA $J$   & $     410 $  & $      61 $  & $     254 $  & $      61 $  & $     432 $  & $      65 $  \\
              UVISTA $H$  & $     432 $  & $      78 $  & $     357 $  & $      77 $  & $     392 $  & $      86 $  \\
   UVISTA $K_\mathrm{S}$  & $     275 $  & $      86 $  & $     263 $  & $      84 $  & $     266 $  & $     110 $  \\
          IRAC $3.6\mu$m  & $     492 $  & $      50 $  & $     589 $  & $      45 $  & $     620 $  & $      68 $  \\
          IRAC $4.5\mu$m  & $     799 $  & $      57 $  & $     729 $  & $      49 $  & $     682 $  & $     108 $  \\
          IRAC $5.8\mu$m  & $     688 $  & $    1702 $  & $    4546 $  & $    2069 $  & $   -1686 $  & $    1819 $  \\
          IRAC $8.0\mu$m  & $    1384 $  & $    2105 $  & $   -2896 $  & $    2461 $  & $    -795 $  & $    2123 $  \\
\enddata
\tablecomments{These measurements are reprocessed fluxes using HSC $z$ band as prior for \texttt{mophongo}.}
\end{deluxetable*}

\section{Sources used to estimate the LF not included in our HST follow-up program}
\label{app:extra}

Here we present the three additional candidate bright $z\sim8$ LBGs, from the sample of \citet{labbe2017},  that we included in our LF estimates (Sect. \ref{sect:LF}). Their selection followed the same methods described in Sect. \ref{sect:selection}. However, due to the lack of HST imaging, we reprocessed their photometry with \texttt{mophongo} adopting the HSC $z$ band as positional and morphological prior: its red effective wavelength together with its depth allowed us to detect almost every source (i.e., neighbouring, potentially contaminating objects) on the UltraVISTA and IRAC mosaics, while its narrow PSF (the narrowest among the ground-based and IRAC data sets) ensures we can consistently use it as prior with \texttt{mophongo} to recover the flux in all the bands.

The fact that our targets are $Y$ or $J$ dropouts, undetected by construction in the $z$ band, does not constitute a major problem for our photometry. Indeed, \texttt{mophongo} can perform the aperture photometry blindly, i.e., without the need of detecting the source of interest.

\begin{figure*}
\includegraphics[width=18cm]{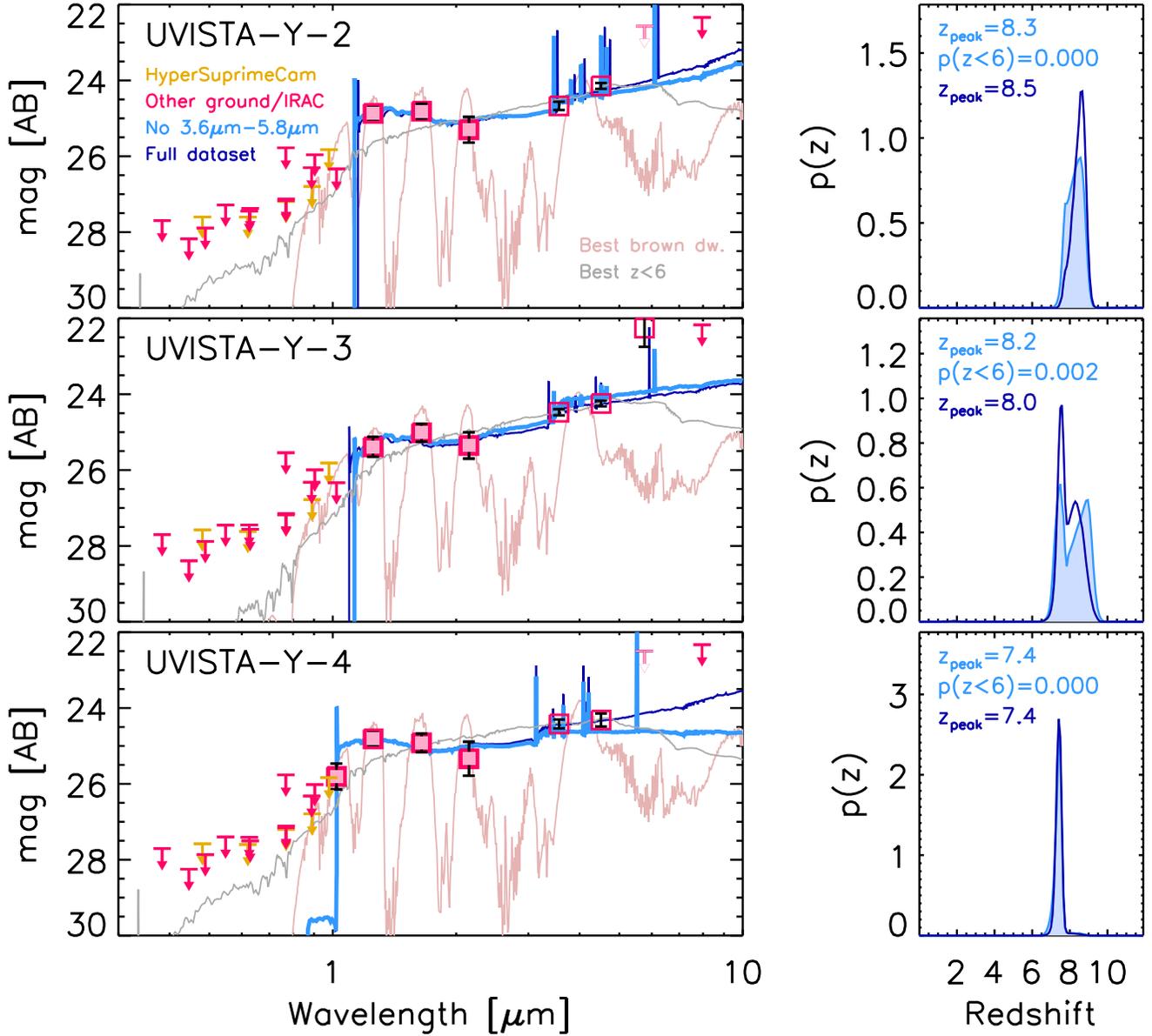}
\caption{{\bf Left panels:} SEDs for the three additional candidate LBGs at $z\gtrsim 8$, included in the estimate of the LF. The colored squares with black errorbars mark the photometric measurements, while arrows represent $2 \sigma$ upper limits. Open squares and arrows mark the IRAC $3.6\mu$m, $4.5\mu$m and $5.8\mu$m bands, not used for the measurement of the fiducial photometric redshift. Photometry in the HyperSuprimeCam Survey bands is represented in yellow. The fiducial best fit SED template from EAzY is indicated by the thick blue curve; the thin dark blue curve represents the best-fit SED when all bands are used for the photometric redshift measurement. Also shown are the best-fitting brown dwarf template (light brown curve) and the solution when the redshift is forced to be $z<6$ (grey curve).   {\bf Right panels:}  Redshift likelihood distributions ($p(z)$) for the three LBGs for the fiducial solution (blue) and for the solution obtained considering the full set of flux measurements. The label in the top-left corner indicates the estimated photometric redshifts. The $p(z)$ are peaked, with no integrated probability for a secondary solution at lower redshifts.  \label{fig:extra}}
\end{figure*}

In Figure \ref{fig:cuts_extra} we present image cutouts in the optical, NIR and IRAC $3.6\mu$m and $4.5\mu$m bands. Table \ref{tab:phot_extra} presents the flux density measurements of these three objects, while Table \ref{tab:obs_extra} lists their main observational properties obtained following the same analysis adopted for our main sample. Their observed and best-fit SEDs and $p(z)$ are shown in Figure \ref{fig:extra}. The $\chi^2$ for the $z\sim8$ solutions are $\chi^2=17.6, 28.7, 16.6$, for UVISTA-Y-2, UVISTA-Y-3 and UVISTA-Y-4, respectively.   The sources are characterized by blue UV continuum slopes ($\beta\sim-2$), excluding red/dusty interloper solutions. Indeed, forcing $z<6$ generates solutions with $\chi^2=76.7, 56.7, 50.1$.  Our measured photometry allows us to exclude a solution as brown dwarves as well ($\chi^2=70.6, 76.5, 57.0$).

Their relative brightness ($H\sim 25$~mag, then, translates into high UV luminosities, with absolute magnitudes $M_\mathrm{UV}\lesssim-22$~mag.

\begin{deluxetable*}{ccccccccc}
\tablecaption{Main properties of the three additional bright candidate $z\sim8$ LBG included in the LF measurement.  \label{tab:obs_extra}}
\tablehead{
\colhead{ID} & \colhead{R.A.} & \colhead{Dec.} & \colhead{H} & \colhead{$z_\mathrm{photo}$\tablenotemark{a}} & \colhead{$p(z>6)$}\tablenotemark{b} & \colhead{$M_\mathrm{UV}$\tablenotemark{c}} & \colhead{$\beta$\tablenotemark{d}} & \colhead{$EW_0(H\beta+\mathrm{[OIII]})$\tablenotemark{e}} \\
 & [J2000] & [J2000] & [mag] & & & [mag] & & [\AA]
}
\startdata
   UVISTA-Y-2 &     10:02:12.558 &       2:30:45.71 & $24.8$ & $ 8.30^{+0.47}_{-0.50}$ & $1.00$ & $ -22.37 \pm  0.20$ & $ -2.45 \pm 0.57$ & $   705^{+ 352}_{- 287} $ \\
   UVISTA-Y-3 &     10:00:32.322 &       1:44:31.26 & $25.0$ & $ 8.24^{+0.81}_{-0.86}$ & $1.00$ & $ -22.02 \pm  0.23$ & $ -1.93 \pm 0.74$ & $   269^{+ 219}_{- 188} $ \\
   UVISTA-Y-4 &     10:00:58.485 &       1:49:55.96 & $24.9$ & $ 7.42^{+0.20}_{-0.23}$ & $1.00$ & $ -22.18 \pm  0.24$ & $ -2.69 \pm 0.68$ & $   124^{+ 412}_{- 331} $ \\
\enddata
\tablenotetext{a}{Best photometric redshift  estimate from EAzY, excluding the IRAC bands from the fit,  and corresponding 68\% confidence interval.}
\tablenotetext{b}{Probability, computed by EAzY, that the solution is at $z>6$.}
\tablenotetext{c}{Absolute magnitudes at rest-frame 1600\AA~from EAzY.}
\tablenotetext{d}{Rest-frame UV continuum slopes from the UltraVISTA $J$, $H$ and $K_\mathrm{S}$ bands.}
\tablenotetext{e}{Rest-frame equivalent width of H$\beta+\mathrm{[OIII]}$ obtained from the $[3.6]-[4.5]$ color assuming an SED flat in $f_\nu$ (i.e. $\beta=-2$). }
\end{deluxetable*}

\bibliographystyle{apj}

\begin{thebibliography}{}
\expandafter\ifx\csname natexlab\endcsname\relax\def\natexlab#1{#1}\fi

\bibitem[{{Aihara} {et~al.}(2017{\natexlab{a}}){Aihara}, {Armstrong},
  {Bickerton}, {Bosch}, {Coupon}, {Furusawa}, {Hayashi}, {Ikeda}, {Kamata},
  {Karoji}, {Kawanomoto}, {Koike}, {Komiyama}, {Lupton}, {Mineo}, {Miyatake},
  {Miyazaki}, {Morokuma}, {Obuchi}, {Oishi}, {Okura}, {Price}, {Takata},
  {Tanaka}, {Tanaka}, {Tanaka}, {Uchida}, {Uraguchi}, {Utsumi}, {Wang},
  {Yamada}, {Yamanoi}, {Yasuda}, {Arimoto}, {Chiba}, {Finet}, {Fujimori},
  {Fujimoto}, {Furusawa}, {Goto}, {Goulding}, {Gunn}, {Harikane}, {Hattori},
  {Hayashi}, {Helminiak}, {Higuchi}, {Hikage}, {Ho}, {Hsieh}, {Huang}, {Huang},
  {Imanishi}, {Iwata}, {Jaelani}, {Jian}, {Kashikawa}, {Katayama}, {Kojima},
  {Konno}, {Koshida}, {Leauthaud}, {Lee}, {Lin}, {Lin}, {Mandelbaum},
  {Matsuoka}, {Medezinski}, {Miyama}, {Momose}, {More}, {More}, {Mukae},
  {Murata}, {Murayama}, {Nagao}, {Nakata}, {Niikura}, {Nishizawa}, {Oguri},
  {Okabe}, {Ono}, {Onodera}, {Onoue}, {Ouchi}, {Pyo}, {Shibuya}, {Shimasaku},
  {Simet}, {Speagle}, {Spergel}, {Strauss}, {Sugahara}, {Sugiyama}, {Suto},
  {Suzuki}, {Tait}, {Takada}, {Terai}, {Toba}, {Turner}, {Uchiyama}, {Umetsu},
  {Urata}, {Usuda}, {Yeh}, \& {Yuma}}]{aihara2017a}
{Aihara}, H., {Armstrong}, R., {Bickerton}, S., {et~al.} 2017{\natexlab{a}},
  ArXiv e-prints, arXiv:1702.08449

\bibitem[{{Aihara} {et~al.}(2017{\natexlab{b}}){Aihara}, {Arimoto},
  {Armstrong}, {Arnouts}, {Bahcall}, {Bickerton}, {Bosch}, {Bundy}, {Capak},
  {Chan}, {Chiba}, {Coupon}, {Egami}, {Enoki}, {Finet}, {Fujimori}, {Fujimoto},
  {Furusawa}, {Furusawa}, {Goto}, {Goulding}, {Greco}, {Greene}, {Gunn},
  {Hamana}, {Harikane}, {Hashimoto}, {Hattori}, {Hayashi}, {Hayashi},
  {He{\l}miniak}, {Higuchi}, {Hikage}, {Ho}, {Hsieh}, {Huang}, {Huang},
  {Ikeda}, {Imanishi}, {Inoue}, {Iwasawa}, {Iwata}, {Jaelani}, {Jian},
  {Kamata}, {Karoji}, {Kashikawa}, {Katayama}, {Kawanomoto}, {Kayo}, {Koda},
  {Koike}, {Kojima}, {Komiyama}, {Konno}, {Koshida}, {Koyama}, {Kusakabe},
  {Leauthaud}, {Lee}, {Lin}, {Lin}, {Lupton}, {Mandelbaum}, {Matsuoka},
  {Medezinski}, {Mineo}, {Miyama}, {Miyatake}, {Miyazaki}, {Momose}, {More},
  {More}, {Moritani}, {Moriya}, {Morokuma}, {Mukae}, {Murata}, {Murayama},
  {Nagao}, {Nakata}, {Niida}, {Niikura}, {Nishizawa}, {Obuchi}, {Oguri},
  {Oishi}, {Okabe}, {Okura}, {Ono}, {Onodera}, {Onoue}, {Osato}, {Ouchi},
  {Price}, {Pyo}, {Sako}, {Okamoto}, {Sawicki}, {Shibuya}, {Shimasaku},
  {Shimono}, {Shirasaki}, {Silverman}, {Simet}, {Speagle}, {Spergel},
  {Strauss}, {Sugahara}, {Sugiyama}, {Suto}, {Suyu}, {Suzuki}, {Tait},
  {Takata}, {Takada}, {Tamura}, {Tanaka}, {Tanaka}, {Tanaka}, {Tanaka},
  {Terai}, {Terashima}, {Toba}, {Toshikawa}, {Turner}, {Uchida}, {Uchiyama},
  {Umetsu}, {Uraguchi}, {Urata}, {Usuda}, {Utsumi}, {Wang}, {Wang}, {Wong},
  {Yabe}, {Yamada}, {Yamanoi}, {Yasuda}, {Yeh}, {Yonehara}, \&
  {Yuma}}]{aihara2017b}
{Aihara}, H., {Arimoto}, N., {Armstrong}, R., {et~al.} 2017{\natexlab{b}},
  ArXiv e-prints, arXiv:1704.05858

\bibitem[{{Ashby} {et~al.}(2013){Ashby}, {Willner}, {Fazio}, {Huang}, {Arendt},
  {Barmby}, {Barro}, {Bell}, {Bouwens}, {Cattaneo}, {Croton}, {Dav{\'e}},
  {Dunlop}, {Egami}, {Faber}, {Finlator}, {Grogin}, {Guhathakurta},
  {Hernquist}, {Hora}, {Illingworth}, {Kashlinsky}, {Koekemoer}, {Koo},
  {Labb{\'e}}, {Li}, {Lin}, {Moseley}, {Nandra}, {Newman}, {Noeske}, {Ouchi},
  {Peth}, {Rigopoulou}, {Robertson}, {Sarajedini}, {Simard}, {Smith}, {Wang},
  {Wechsler}, {Weiner}, {Wilson}, {Wuyts}, {Yamada}, \& {Yan}}]{ashby2013}
{Ashby}, M.~L.~N., {Willner}, S.~P., {Fazio}, G.~G., {et~al.} 2013, \apj, 769,
  80

\bibitem[{{Ashby} {et~al.}(2015){Ashby}, {Willner}, {Fazio}, {Dunlop}, {Egami},
  {Faber}, {Ferguson}, {Grogin}, {Hora}, {Huang}, {Koekemoer}, {Labb{\'e}}, \&
  {Wang}}]{ashby2015}
---. 2015, \apjs, 218, 33

\bibitem[{{Ashby et al.}(2017, in preparation)}]{ashby2017}
{Ashby et al.} 2017, in preparation, ApJS

\bibitem[{{Avni} \& {Bahcall}(1980)}]{avni1980}
{Avni}, Y., \& {Bahcall}, J.~N. 1980, \apj, 235, 694

\bibitem[{{Beckwith} {et~al.}(2006){Beckwith}, {Stiavelli}, {Koekemoer},
  {Caldwell}, {Ferguson}, {Hook}, {Lucas}, {Bergeron}, {Corbin}, {Jogee},
  {Panagia}, {Robberto}, {Royle}, {Somerville}, \& {Sosey}}]{beckwith2006}
{Beckwith}, S.~V.~W., {Stiavelli}, M., {Koekemoer}, A.~M., {et~al.} 2006, \aj,
  132, 1729

\bibitem[{{Bernard} {et~al.}(2016){Bernard}, {Carrasco}, {Trenti}, {Oesch},
  {Wu}, {Bradley}, {Schmidt}, {Bouwens}, {Calvi}, {Mason}, {Stiavelli}, \&
  {Treu}}]{bernard2016}
{Bernard}, S.~R., {Carrasco}, D., {Trenti}, M., {et~al.} 2016, \apj, 827, 76

\bibitem[{{Bertin} \& {Arnouts}(1996)}]{bertin1996}
{Bertin}, E., \& {Arnouts}, S. 1996, \aaps, 117, 393

\bibitem[{{Bouwens} {et~al.}(2016{\natexlab{a}}){Bouwens}, {Oesch},
  {Illingworth}, {Ellis}, \& {Stefanon}}]{bouwens2017}
{Bouwens}, R.~J., {Oesch}, P.~A., {Illingworth}, G.~D., {Ellis}, R.~S., \&
  {Stefanon}, M. 2016{\natexlab{a}}, ArXiv e-prints, arXiv:1610.00283

\bibitem[{{Bouwens} {et~al.}(2011){Bouwens}, {Illingworth}, {Oesch},
  {Labb{\'e}}, {Trenti}, {van Dokkum}, {Franx}, {Stiavelli}, {Carollo},
  {Magee}, \& {Gonzalez}}]{bouwens2011}
{Bouwens}, R.~J., {Illingworth}, G.~D., {Oesch}, P.~A., {et~al.} 2011, \apj,
  737, 90

\bibitem[{{Bouwens} {et~al.}(2012){Bouwens}, {Illingworth}, {Oesch}, {Franx},
  {Labb{\'e}}, {Trenti}, {van Dokkum}, {Carollo}, {Gonz{\'a}lez}, {Smit}, \&
  {Magee}}]{bouwens2012}
---. 2012, \apj, 754, 83

\bibitem[{{Bouwens} {et~al.}(2014){Bouwens}, {Bradley}, {Zitrin}, {Coe},
  {Franx}, {Zheng}, {Smit}, {Host}, {Postman}, {Moustakas}, {Labb{\'e}},
  {Carrasco}, {Molino}, {Donahue}, {Kelson}, {Meneghetti}, {Ben{\'{\i}}tez},
  {Lemze}, {Umetsu}, {Broadhurst}, {Moustakas}, {Rosati}, {Jouvel},
  {Bartelmann}, {Ford}, {Graves}, {Grillo}, {Infante}, {Jimenez-Teja}, {Lahav},
  {Maoz}, {Medezinski}, {Melchior}, {Merten}, {Nonino}, {Ogaz}, \&
  {Seitz}}]{bouwens2014}
{Bouwens}, R.~J., {Bradley}, L., {Zitrin}, A., {et~al.} 2014, \apj, 795, 126

\bibitem[{{Bouwens} {et~al.}(2015){Bouwens}, {Illingworth}, {Oesch}, {Trenti},
  {Labb{\'e}}, {Bradley}, {Carollo}, {van Dokkum}, {Gonzalez}, {Holwerda},
  {Franx}, {Spitler}, {Smit}, \& {Magee}}]{bouwens2015}
{Bouwens}, R.~J., {Illingworth}, G.~D., {Oesch}, P.~A., {et~al.} 2015, \apj,
  803, 34

\bibitem[{{Bouwens} {et~al.}(2016{\natexlab{b}}){Bouwens}, {Oesch},
  {Labb{\'e}}, {Illingworth}, {Fazio}, {Coe}, {Holwerda}, {Smit}, {Stefanon},
  {van Dokkum}, {Trenti}, {Ashby}, {Huang}, {Spitler}, {Straatman}, {Bradley},
  \& {Magee}}]{bouwens2016}
{Bouwens}, R.~J., {Oesch}, P.~A., {Labb{\'e}}, I., {et~al.} 2016{\natexlab{b}},
  \apj, 830, 67

\bibitem[{{Bowler} {et~al.}(2017){Bowler}, {Dunlop}, {McLure}, \&
  {McLeod}}]{bowler2017}
{Bowler}, R.~A.~A., {Dunlop}, J.~S., {McLure}, R.~J., \& {McLeod}, D.~J. 2017,
  \mnras, 466, 3612

\bibitem[{{Bowler} {et~al.}(2014){Bowler}, {Dunlop}, {McLure}, {Rogers},
  {McCracken}, {Milvang-Jensen}, {Furusawa}, {Fynbo}, {Taniguchi}, {Afonso},
  {Bremer}, \& {Le F{\`e}vre}}]{bowler2014}
{Bowler}, R.~A.~A., {Dunlop}, J.~S., {McLure}, R.~J., {et~al.} 2014, \mnras,
  440, 2810

\bibitem[{{Bowler} {et~al.}(2015){Bowler}, {Dunlop}, {McLure}, {McCracken},
  {Milvang-Jensen}, {Furusawa}, {Taniguchi}, {Le F{\`e}vre}, {Fynbo}, {Jarvis},
  \& {H{\"a}u{\ss}ler}}]{bowler2015}
---. 2015, \mnras, 452, 1817

\bibitem[{{Bradley} {et~al.}(2012){Bradley}, {Trenti}, {Oesch}, {Stiavelli},
  {Treu}, {Bouwens}, {Shull}, {Holwerda}, \& {Pirzkal}}]{bradley2012}
{Bradley}, L.~D., {Trenti}, M., {Oesch}, P.~A., {et~al.} 2012, \apj, 760, 108

\bibitem[{{Brammer} {et~al.}(2008){Brammer}, {van Dokkum}, \&
  {Coppi}}]{brammer2008}
{Brammer}, G.~B., {van Dokkum}, P.~G., \& {Coppi}, P. 2008, \apj, 686, 1503

\bibitem[{{Brammer} {et~al.}(2012){Brammer}, {van Dokkum}, {Franx},
  {Fumagalli}, {Patel}, {Rix}, {Skelton}, {Kriek}, {Nelson}, {Schmidt},
  {Bezanson}, {da Cunha}, {Erb}, {Fan}, {F{\"o}rster Schreiber}, {Illingworth},
  {Labb{\'e}}, {Leja}, {Lundgren}, {Magee}, {Marchesini}, {McCarthy},
  {Momcheva}, {Muzzin}, {Quadri}, {Steidel}, {Tal}, {Wake}, {Whitaker}, \&
  {Williams}}]{brammer2012}
{Brammer}, G.~B., {van Dokkum}, P.~G., {Franx}, M., {et~al.} 2012, \apjs, 200,
  13

\bibitem[{{Bruzual} \& {Charlot}(2003)}]{bruzual2003}
{Bruzual}, G., \& {Charlot}, S. 2003, \mnras, 344, 1000

\bibitem[{{Burgasser}(2014)}]{burgasser2014}
{Burgasser}, A.~J. 2014, in Astronomical Society of India Conference Series,
  Vol.~11, Astronomical Society of India Conference Series

\bibitem[{{Burrows} {et~al.}(2006){Burrows}, {Sudarsky}, \&
  {Hubeny}}]{burrows2006}
{Burrows}, A., {Sudarsky}, D., \& {Hubeny}, I. 2006, \apj, 640, 1063

\bibitem[{{Calvi} {et~al.}(2016){Calvi}, {Trenti}, {Stiavelli}, {Oesch},
  {Bradley}, {Schmidt}, {Coe}, {Brammer}, {Bernard}, {Bouwens}, {Carrasco},
  {Carollo}, {Holwerda}, {MacKenty}, {Mason}, {Shull}, \& {Treu}}]{calvi2016}
{Calvi}, V., {Trenti}, M., {Stiavelli}, M., {et~al.} 2016, \apj, 817, 120

\bibitem[{{Calzetti} {et~al.}(2000){Calzetti}, {Armus}, {Bohlin}, {Kinney},
  {Koornneef}, \& {Storchi-Bergmann}}]{calzetti2000}
{Calzetti}, D., {Armus}, L., {Bohlin}, R.~C., {et~al.} 2000, \apj, 533, 682

\bibitem[{{Cappelluti} {et~al.}(2009){Cappelluti}, {Brusa}, {Hasinger},
  {Comastri}, {Zamorani}, {Finoguenov}, {Gilli}, {Puccetti}, {Miyaji},
  {Salvato}, {Vignali}, {Aldcroft}, {B{\"o}hringer}, {Brunner}, {Civano},
  {Elvis}, {Fiore}, {Fruscione}, {Griffiths}, {Guzzo}, {Iovino}, {Koekemoer},
  {Mainieri}, {Scoville}, {Shopbell}, {Silverman}, \& {Urry}}]{cappelluti2009}
{Cappelluti}, N., {Brusa}, M., {Hasinger}, G., {et~al.} 2009, \aap, 497, 635

\bibitem[{{Caputi} {et~al.}(2017){Caputi}, {Deshmukh}, {Ashby}, {Cowley},
  {Bisigello}, {Fazio}, {Fynbo}, {Le Fevre}, {Milvang-Jensen}, \&
  {Ilbert}}]{caputi2017}
{Caputi}, K.~I., {Deshmukh}, S., {Ashby}, M.~L.~N., {et~al.} 2017, ArXiv
  e-prints, arXiv:1705.06179

\bibitem[{{Castellano} {et~al.}(2012){Castellano}, {Fontana}, {Grazian},
  {Pentericci}, {Santini}, {Koekemoer}, {Cristiani}, {Galametz}, {Gallerani},
  {Vanzella}, {Boutsia}, {Gallozzi}, {Giallongo}, {Maiolino}, {Menci}, \&
  {Paris}}]{castellano2012}
{Castellano}, M., {Fontana}, A., {Grazian}, A., {et~al.} 2012, \aap, 540, A39

\bibitem[{{Civano} {et~al.}(2016){Civano}, {Marchesi}, {Comastri}, {Urry},
  {Elvis}, {Cappelluti}, {Puccetti}, {Brusa}, {Zamorani}, {Hasinger},
  {Aldcroft}, {Alexander}, {Allevato}, {Brunner}, {Capak}, {Finoguenov},
  {Fiore}, {Fruscione}, {Gilli}, {Glotfelty}, {Griffiths}, {Hao}, {Harrison},
  {Jahnke}, {Kartaltepe}, {Karim}, {LaMassa}, {Lanzuisi}, {Miyaji}, {Ranalli},
  {Salvato}, {Sargent}, {Scoville}, {Schawinski}, {Schinnerer}, {Silverman},
  {Smolcic}, {Stern}, {Toft}, {Trakhenbrot}, {Treister}, \&
  {Vignali}}]{civano2016}
{Civano}, F., {Marchesi}, S., {Comastri}, A., {et~al.} 2016, \apj, 819, 62

\bibitem[{{Curtis-Lake} {et~al.}(2016){Curtis-Lake}, {McLure}, {Dunlop},
  {Rogers}, {Targett}, {Dekel}, {Ellis}, {Faber}, {Ferguson}, {Grogin},
  {Kocevski}, {Koekemoer}, {Lai}, {M{\'a}rmol-Queralt{\'o}}, \&
  {Robertson}}]{curtis-lake2016}
{Curtis-Lake}, E., {McLure}, R.~J., {Dunlop}, J.~S., {et~al.} 2016, \mnras,
  457, 440

\bibitem[{{Duncan} \& {Conselice}(2015)}]{duncan2015}
{Duncan}, K., \& {Conselice}, C.~J. 2015, \mnras, 451, 2030

\bibitem[{{Duncan} {et~al.}(2014){Duncan}, {Conselice}, {Mortlock}, {Hartley},
  {Guo}, {Ferguson}, {Dav{\'e}}, {Lu}, {Ownsworth}, {Ashby}, {Dekel},
  {Dickinson}, {Faber}, {Giavalisco}, {Grogin}, {Kocevski}, {Koekemoer},
  {Somerville}, \& {White}}]{duncan2014}
{Duncan}, K., {Conselice}, C.~J., {Mortlock}, A., {et~al.} 2014, \mnras, 444,
  2960

\bibitem[{{Dunlop} {et~al.}(2013){Dunlop}, {Rogers}, {McLure}, {Ellis},
  {Robertson}, {Koekemoer}, {Dayal}, {Curtis-Lake}, {Wild}, {Charlot},
  {Bowler}, {Schenker}, {Ouchi}, {Ono}, {Cirasuolo}, {Furlanetto}, {Stark},
  {Targett}, \& {Schneider}}]{dunlop2013}
{Dunlop}, J.~S., {Rogers}, A.~B., {McLure}, R.~J., {et~al.} 2013, \mnras, 432,
  3520

\bibitem[{{Erb} {et~al.}(2006){Erb}, {Steidel}, {Shapley}, {Pettini}, {Reddy},
  \& {Adelberger}}]{erb2006}
{Erb}, D.~K., {Steidel}, C.~C., {Shapley}, A.~E., {et~al.} 2006, \apj, 647, 128

\bibitem[{{Erben} {et~al.}(2009){Erben}, {Hildebrandt}, {Lerchster}, {Hudelot},
  {Benjamin}, {van Waerbeke}, {Schrabback}, {Brimioulle}, {Cordes}, {Dietrich},
  {Holhjem}, {Schirmer}, \& {Schneider}}]{erben2009}
{Erben}, T., {Hildebrandt}, H., {Lerchster}, M., {et~al.} 2009, \aap, 493, 1197

\bibitem[{{Fern{\'a}ndez-Soto} {et~al.}(1999){Fern{\'a}ndez-Soto}, {Lanzetta},
  \& {Yahil}}]{fernandez-soto1999}
{Fern{\'a}ndez-Soto}, A., {Lanzetta}, K.~M., \& {Yahil}, A. 1999, \apj, 513, 34

\bibitem[{{Finkelstein} {et~al.}(2012){Finkelstein}, {Papovich}, {Salmon},
  {Finlator}, {Dickinson}, {Ferguson}, {Giavalisco}, {Koekemoer}, {Reddy},
  {Bassett}, {Conselice}, {Dunlop}, {Faber}, {Grogin}, {Hathi}, {Kocevski},
  {Lai}, {Lee}, {McLure}, {Mobasher}, \& {Newman}}]{finkelstein2012}
{Finkelstein}, S.~L., {Papovich}, C., {Salmon}, B., {et~al.} 2012, \apj, 756,
  164

\bibitem[{{Finkelstein} {et~al.}(2015){Finkelstein}, {Ryan}, {Papovich},
  {Dickinson}, {Song}, {Somerville}, {Ferguson}, {Salmon}, {Giavalisco},
  {Koekemoer}, {Ashby}, {Behroozi}, {Castellano}, {Dunlop}, {Faber}, {Fazio},
  {Fontana}, {Grogin}, {Hathi}, {Jaacks}, {Kocevski}, {Livermore}, {McLure},
  {Merlin}, {Mobasher}, {Newman}, {Rafelski}, {Tilvi}, \&
  {Willner}}]{finkelstein2015a}
{Finkelstein}, S.~L., {Ryan}, Jr., R.~E., {Papovich}, C., {et~al.} 2015, \apj,
  810, 71

\bibitem[{{Gehrels}(1986)}]{gehrels1986}
{Gehrels}, N. 1986, \apj, 303, 336

\bibitem[{{Grogin} {et~al.}(2011){Grogin}, {Kocevski}, {Faber}, {Ferguson},
  {Koekemoer}, {Riess}, {Acquaviva}, {Alexander}, {Almaini}, {Ashby}, {Barden},
  {Bell}, {Bournaud}, {Brown}, {Caputi}, {Casertano}, {Cassata}, {Castellano},
  {Challis}, {Chary}, {Cheung}, {Cirasuolo}, {Conselice}, {Roshan Cooray},
  {Croton}, {Daddi}, {Dahlen}, {Dav{\'e}}, {de Mello}, {Dekel}, {Dickinson},
  {Dolch}, {Donley}, {Dunlop}, {Dutton}, {Elbaz}, {Fazio}, {Filippenko},
  {Finkelstein}, {Fontana}, {Gardner}, {Garnavich}, {Gawiser}, {Giavalisco},
  {Grazian}, {Guo}, {Hathi}, {H{\"a}ussler}, {Hopkins}, {Huang}, {Huang},
  {Jha}, {Kartaltepe}, {Kirshner}, {Koo}, {Lai}, {Lee}, {Li}, {Lotz}, {Lucas},
  {Madau}, {McCarthy}, {McGrath}, {McIntosh}, {McLure}, {Mobasher},
  {Moustakas}, {Mozena}, {Nandra}, {Newman}, {Niemi}, {Noeske}, {Papovich},
  {Pentericci}, {Pope}, {Primack}, {Rajan}, {Ravindranath}, {Reddy}, {Renzini},
  {Rix}, {Robaina}, {Rodney}, {Rosario}, {Rosati}, {Salimbeni}, {Scarlata},
  {Siana}, {Simard}, {Smidt}, {Somerville}, {Spinrad}, {Straughn}, {Strolger},
  {Telford}, {Teplitz}, {Trump}, {van der Wel}, {Villforth}, {Wechsler},
  {Weiner}, {Wiklind}, {Wild}, {Wilson}, {Wuyts}, {Yan}, \& {Yun}}]{grogin2011}
{Grogin}, N.~A., {Kocevski}, D.~D., {Faber}, S.~M., {et~al.} 2011, \apjs, 197,
  35

\bibitem[{{Hildebrandt} {et~al.}(2009){Hildebrandt}, {Pielorz}, {Erben}, {van
  Waerbeke}, {Simon}, \& {Capak}}]{hildebrandt2009}
{Hildebrandt}, H., {Pielorz}, J., {Erben}, T., {et~al.} 2009, \aap, 498, 725

\bibitem[{{Holwerda} {et~al.}(2015){Holwerda}, {Bouwens}, {Oesch}, {Smit},
  {Illingworth}, \& {Labbe}}]{holwerda2015}
{Holwerda}, B.~W., {Bouwens}, R., {Oesch}, P., {et~al.} 2015, \apj, 808, 6

\bibitem[{{Holwerda} {et~al.}(2014){Holwerda}, {Trenti}, {Clarkson}, {Sahu},
  {Bradley}, {Stiavelli}, {Pirzkal}, {De Marchi}, {Andersen}, {Bouwens}, \&
  {Ryan}}]{holwerda2014}
{Holwerda}, B.~W., {Trenti}, M., {Clarkson}, W., {et~al.} 2014, \apj, 788, 77

\bibitem[{{Huang} {et~al.}(2013){Huang}, {Ferguson}, {Ravindranath}, \&
  {Su}}]{huang2013}
{Huang}, K.-H., {Ferguson}, H.~C., {Ravindranath}, S., \& {Su}, J. 2013, \apj,
  765, 68

\bibitem[{{Huertas-Company} {et~al.}(2015){Huertas-Company},
  {P{\'e}rez-Gonz{\'a}lez}, {Mei}, {Shankar}, {Bernardi}, {Daddi}, {Barro},
  {Cabrera-Vives}, {Cattaneo}, {Dimauro}, \& {Gravet}}]{huertas-company2015}
{Huertas-Company}, M., {P{\'e}rez-Gonz{\'a}lez}, P.~G., {Mei}, S., {et~al.}
  2015, \apj, 809, 95

\bibitem[{{Illingworth} {et~al.}(2013){Illingworth}, {Magee}, {Oesch},
  {Bouwens}, {Labb{\'e}}, {Stiavelli}, {van Dokkum}, {Franx}, {Trenti},
  {Carollo}, \& {Gonzalez}}]{illingworth2013}
{Illingworth}, G.~D., {Magee}, D., {Oesch}, P.~A., {et~al.} 2013, \apjs, 209, 6

\bibitem[{{Ishigaki} {et~al.}(2017){Ishigaki}, {Kawamata}, {Ouchi}, {Oguri}, \&
  {Shimasaku}}]{ishigaki2017}
{Ishigaki}, M., {Kawamata}, R., {Ouchi}, M., {Oguri}, M., \& {Shimasaku}, K.
  2017, ArXiv e-prints, arXiv:1702.04867

\bibitem[{{Jiang} {et~al.}(2013){Jiang}, {Egami}, {Mechtley}, {Fan}, {Cohen},
  {Windhorst}, {Dav{\'e}}, {Finlator}, {Kashikawa}, {Ouchi}, \&
  {Shimasaku}}]{jiang2013}
{Jiang}, L., {Egami}, E., {Mechtley}, M., {et~al.} 2013, \apj, 772, 99

\bibitem[{{Kennicutt}(1998)}]{kennicutt1998}
{Kennicutt}, Jr., R.~C. 1998, \araa, 36, 189

\bibitem[{{Koekemoer} {et~al.}(2003){Koekemoer}, {Fruchter}, {Hook}, \&
  {Hack}}]{koekemoer2003}
{Koekemoer}, A.~M., {Fruchter}, A.~S., {Hook}, R.~N., \& {Hack}, W. 2003, in
  HST Calibration Workshop : Hubble after the Installation of the ACS and the
  NICMOS Cooling System, ed. S.~{Arribas}, A.~{Koekemoer}, \& B.~{Whitmore},
  337

\bibitem[{{Koekemoer} {et~al.}(2011){Koekemoer}, {Faber}, {Ferguson}, {Grogin},
  {Kocevski}, {Koo}, {Lai}, {Lotz}, {Lucas}, {McGrath}, {Ogaz}, {Rajan},
  {Riess}, {Rodney}, {Strolger}, {Casertano}, {Castellano}, {Dahlen},
  {Dickinson}, {Dolch}, {Fontana}, {Giavalisco}, {Grazian}, {Guo}, {Hathi},
  {Huang}, {van der Wel}, {Yan}, {Acquaviva}, {Alexander}, {Almaini}, {Ashby},
  {Barden}, {Bell}, {Bournaud}, {Brown}, {Caputi}, {Cassata}, {Challis},
  {Chary}, {Cheung}, {Cirasuolo}, {Conselice}, {Roshan Cooray}, {Croton},
  {Daddi}, {Dav{\'e}}, {de Mello}, {de Ravel}, {Dekel}, {Donley}, {Dunlop},
  {Dutton}, {Elbaz}, {Fazio}, {Filippenko}, {Finkelstein}, {Frazer}, {Gardner},
  {Garnavich}, {Gawiser}, {Gruetzbauch}, {Hartley}, {H{\"a}ussler},
  {Herrington}, {Hopkins}, {Huang}, {Jha}, {Johnson}, {Kartaltepe},
  {Khostovan}, {Kirshner}, {Lani}, {Lee}, {Li}, {Madau}, {McCarthy},
  {McIntosh}, {McLure}, {McPartland}, {Mobasher}, {Moreira}, {Mortlock},
  {Moustakas}, {Mozena}, {Nandra}, {Newman}, {Nielsen}, {Niemi}, {Noeske},
  {Papovich}, {Pentericci}, {Pope}, {Primack}, {Ravindranath}, {Reddy},
  {Renzini}, {Rix}, {Robaina}, {Rosario}, {Rosati}, {Salimbeni}, {Scarlata},
  {Siana}, {Simard}, {Smidt}, {Snyder}, {Somerville}, {Spinrad}, {Straughn},
  {Telford}, {Teplitz}, {Trump}, {Vargas}, {Villforth}, {Wagner}, {Wandro},
  {Wechsler}, {Weiner}, {Wiklind}, {Wild}, {Wilson}, {Wuyts}, \&
  {Yun}}]{koekemoer2011}
{Koekemoer}, A.~M., {Faber}, S.~M., {Ferguson}, H.~C., {et~al.} 2011, \apjs,
  197, 36

\bibitem[{{Labb{\'e}} {et~al.}(2006){Labb{\'e}}, {Bouwens}, {Illingworth}, \&
  {Franx}}]{labbe2006}
{Labb{\'e}}, I., {Bouwens}, R., {Illingworth}, G.~D., \& {Franx}, M. 2006,
  \apjl, 649, L67

\bibitem[{{Labb{\'e}} {et~al.}(2010{\natexlab{a}}){Labb{\'e}}, {Gonz{\'a}lez},
  {Bouwens}, {Illingworth}, {Franx}, {Trenti}, {Oesch}, {van Dokkum},
  {Stiavelli}, {Carollo}, {Kriek}, \& {Magee}}]{labbe2010a}
{Labb{\'e}}, I., {Gonz{\'a}lez}, V., {Bouwens}, R.~J., {et~al.}
  2010{\natexlab{a}}, \apjl, 716, L103

\bibitem[{{Labb{\'e}} {et~al.}(2010{\natexlab{b}}){Labb{\'e}}, {Gonz{\'a}lez},
  {Bouwens}, {Illingworth}, {Oesch}, {van Dokkum}, {Carollo}, {Franx},
  {Stiavelli}, {Trenti}, {Magee}, \& {Kriek}}]{labbe2010b}
---. 2010{\natexlab{b}}, \apjl, 708, L26

\bibitem[{{Labb{\'e}} {et~al.}(2013){Labb{\'e}}, {Oesch}, {Bouwens},
  {Illingworth}, {Magee}, {Gonz{\'a}lez}, {Carollo}, {Franx}, {Trenti}, {van
  Dokkum}, \& {Stiavelli}}]{labbe2013}
{Labb{\'e}}, I., {Oesch}, P.~A., {Bouwens}, R.~J., {et~al.} 2013, \apjl, 777,
  L19

\bibitem[{{Labb{\'e}} {et~al.}(2015){Labb{\'e}}, {Oesch}, {Illingworth}, {van
  Dokkum}, {Bouwens}, {Franx}, {Carollo}, {Trenti}, {Holden}, {Smit},
  {Gonz{\'a}lez}, {Magee}, {Stiavelli}, \& {Stefanon}}]{labbe2015}
{Labb{\'e}}, I., {Oesch}, P.~A., {Illingworth}, G.~D., {et~al.} 2015, \apjs,
  221, 23

\bibitem[{{Labb\'e et al.}(2017, in preparation)}]{labbe2017}
{Labb\'e et al.} 2017, in preparation

\bibitem[{{Laidler} {et~al.}(2007){Laidler}, {Papovich}, {Grogin}, {Idzi},
  {Dickinson}, {Ferguson}, {Hilbert}, {Clubb}, \& {Ravindranath}}]{laidler2007}
{Laidler}, V.~G., {Papovich}, C., {Grogin}, N.~A., {et~al.} 2007, \pasp, 119,
  1325

\bibitem[{{Law} {et~al.}(2007){Law}, {Steidel}, {Erb}, {Pettini}, {Reddy},
  {Shapley}, {Adelberger}, \& {Simenc}}]{law2007}
{Law}, D.~R., {Steidel}, C.~C., {Erb}, D.~K., {et~al.} 2007, \apj, 656, 1

\bibitem[{{Lotz} {et~al.}(2017){Lotz}, {Koekemoer}, {Coe}, {Grogin}, {Capak},
  {Mack}, {Anderson}, {Avila}, {Barker}, {Borncamp}, {Brammer}, {Durbin},
  {Gunning}, {Hilbert}, {Jenkner}, {Khandrika}, {Levay}, {Lucas}, {MacKenty},
  {Ogaz}, {Porterfield}, {Reid}, {Robberto}, {Royle}, {Smith},
  {Storrie-Lombardi}, {Sunnquist}, {Surace}, {Taylor}, {Williams}, {Bullock},
  {Dickinson}, {Finkelstein}, {Natarajan}, {Richard}, {Robertson}, {Tumlinson},
  {Zitrin}, {Flanagan}, {Sembach}, {Soifer}, \& {Mountain}}]{lotz2017}
{Lotz}, J.~M., {Koekemoer}, A., {Coe}, D., {et~al.} 2017, \apj, 837, 97

\bibitem[{{Marchesi} {et~al.}(2016){Marchesi}, {Civano}, {Elvis}, {Salvato},
  {Brusa}, {Comastri}, {Gilli}, {Hasinger}, {Lanzuisi}, {Miyaji}, {Treister},
  {Urry}, {Vignali}, {Zamorani}, {Allevato}, {Cappelluti}, {Cardamone},
  {Finoguenov}, {Griffiths}, {Karim}, {Laigle}, {LaMassa}, {Jahnke}, {Ranalli},
  {Schawinski}, {Schinnerer}, {Silverman}, {Smolcic}, {Suh}, \&
  {Trakhtenbrot}}]{marchesi2016}
{Marchesi}, S., {Civano}, F., {Elvis}, M., {et~al.} 2016, \apj, 817, 34

\bibitem[{{McCracken} {et~al.}(2012){McCracken}, {Milvang-Jensen}, {Dunlop},
  {Franx}, {Fynbo}, {Le F{\`e}vre}, {Holt}, {Caputi}, {Goranova}, {Buitrago},
  {Emerson}, {Freudling}, {Hudelot}, {L{\'o}pez-Sanjuan}, {Magnard}, {Mellier},
  {M{\o}ller}, {Nilsson}, {Sutherland}, {Tasca}, \& {Zabl}}]{mccracken2012}
{McCracken}, H.~J., {Milvang-Jensen}, B., {Dunlop}, J., {et~al.} 2012, \aap,
  544, A156

\bibitem[{{McLeod} {et~al.}(2016){McLeod}, {McLure}, \& {Dunlop}}]{mcleod2016}
{McLeod}, D.~J., {McLure}, R.~J., \& {Dunlop}, J.~S. 2016, \mnras, 459, 3812

\bibitem[{{McLeod} {et~al.}(2015){McLeod}, {McLure}, {Dunlop}, {Robertson},
  {Ellis}, \& {Targett}}]{mcleod2015}
{McLeod}, D.~J., {McLure}, R.~J., {Dunlop}, J.~S., {et~al.} 2015, \mnras, 450,
  3032

\bibitem[{{McLure} {et~al.}(2013){McLure}, {Dunlop}, {Bowler}, {Curtis-Lake},
  {Schenker}, {Ellis}, {Robertson}, {Koekemoer}, {Rogers}, {Ono}, {Ouchi},
  {Charlot}, {Wild}, {Stark}, {Furlanetto}, {Cirasuolo}, \&
  {Targett}}]{mclure2013}
{McLure}, R.~J., {Dunlop}, J.~S., {Bowler}, R.~A.~A., {et~al.} 2013, \mnras,
  432, 2696

\bibitem[{{Merlin} {et~al.}(2015){Merlin}, {Fontana}, {Ferguson}, {Dunlop},
  {Elbaz}, {Bourne}, {Bruce}, {Buitrago}, {Castellano}, {Schreiber}, {Grazian},
  {McLure}, {Okumura}, {Shu}, {Wang}, {Amor{\'{\i}}n}, {Boutsia}, {Cappelluti},
  {Comastri}, {Derriere}, {Faber}, \& {Santini}}]{merlin2015}
{Merlin}, E., {Fontana}, A., {Ferguson}, H.~C., {et~al.} 2015, \aap, 582, A15

\bibitem[{{Momcheva} {et~al.}(2016){Momcheva}, {Brammer}, {van Dokkum},
  {Skelton}, {Whitaker}, {Nelson}, {Fumagalli}, {Maseda}, {Leja}, {Franx},
  {Rix}, {Bezanson}, {Da Cunha}, {Dickey}, {F{\"o}rster Schreiber},
  {Illingworth}, {Kriek}, {Labb{\'e}}, {Ulf Lange}, {Lundgren}, {Magee},
  {Marchesini}, {Oesch}, {Pacifici}, {Patel}, {Price}, {Tal}, {Wake}, {van der
  Wel}, \& {Wuyts}}]{momcheva2016}
{Momcheva}, I.~G., {Brammer}, G.~B., {van Dokkum}, P.~G., {et~al.} 2016, \apjs,
  225, 27

\bibitem[{{Mortlock} {et~al.}(2013){Mortlock}, {Conselice}, {Hartley},
  {Ownsworth}, {Lani}, {Bluck}, {Almaini}, {Duncan}, {van der Wel},
  {Koekemoer}, {Dekel}, {Dav{\'e}}, {Ferguson}, {de Mello}, {Newman}, {Faber},
  {Grogin}, {Kocevski}, \& {Lai}}]{mortlock2013}
{Mortlock}, A., {Conselice}, C.~J., {Hartley}, W.~G., {et~al.} 2013, \mnras,
  433, 1185

\bibitem[{{Moster} {et~al.}(2011){Moster}, {Somerville}, {Newman}, \&
  {Rix}}]{moster2011}
{Moster}, B.~P., {Somerville}, R.~S., {Newman}, J.~A., \& {Rix}, H.-W. 2011,
  \apj, 731, 113

\bibitem[{{Oesch} {et~al.}(2010){Oesch}, {Bouwens}, {Carollo}, {Illingworth},
  {Trenti}, {Stiavelli}, {Magee}, {Labb{\'e}}, \& {Franx}}]{oesch2010}
{Oesch}, P.~A., {Bouwens}, R.~J., {Carollo}, C.~M., {et~al.} 2010, \apjl, 709,
  L21

\bibitem[{{Oesch} {et~al.}(2013){Oesch}, {Labb{\'e}}, {Bouwens}, {Illingworth},
  {Gonzalez}, {Franx}, {Trenti}, {Holden}, {van Dokkum}, \&
  {Magee}}]{oesch2013}
{Oesch}, P.~A., {Labb{\'e}}, I., {Bouwens}, R.~J., {et~al.} 2013, \apj, 772,
  136

\bibitem[{{Oesch} {et~al.}(2014){Oesch}, {Bouwens}, {Illingworth}, {Labb{\'e}},
  {Smit}, {Franx}, {van Dokkum}, {Momcheva}, {Ashby}, {Fazio}, {Huang},
  {Willner}, {Gonzalez}, {Magee}, {Trenti}, {Brammer}, {Skelton}, \&
  {Spitler}}]{oesch2014}
{Oesch}, P.~A., {Bouwens}, R.~J., {Illingworth}, G.~D., {et~al.} 2014, \apj,
  786, 108

\bibitem[{{Oesch} {et~al.}(2016){Oesch}, {Brammer}, {van Dokkum},
  {Illingworth}, {Bouwens}, {Labb{\'e}}, {Franx}, {Momcheva}, {Ashby}, {Fazio},
  {Gonzalez}, {Holden}, {Magee}, {Skelton}, {Smit}, {Spitler}, {Trenti}, \&
  {Willner}}]{oesch2016}
{Oesch}, P.~A., {Brammer}, G., {van Dokkum}, P.~G., {et~al.} 2016, \apj, 819,
  129

\bibitem[{{Ono} {et~al.}(2013){Ono}, {Ouchi}, {Curtis-Lake}, {Schenker},
  {Ellis}, {McLure}, {Dunlop}, {Robertson}, {Koekemoer}, {Bowler}, {Rogers},
  {Schneider}, {Charlot}, {Stark}, {Shimasaku}, {Furlanetto}, \&
  {Cirasuolo}}]{ono2013}
{Ono}, Y., {Ouchi}, M., {Curtis-Lake}, E., {et~al.} 2013, \apj, 777, 155

\bibitem[{{Ono} {et~al.}(2017){Ono}, {Ouchi}, {Harikane}, {Toshikawa}, {Rauch},
  {Yuma}, {Sawicki}, {Shibuya}, {Shimasaku}, {Oguri}, {Willott}, {Akhlaghi},
  {Akiyama}, {Coupon}, {Kashikawa}, {Komiyama}, {Konno}, {Lin}, {Matsuoka},
  {Miyazaki}, {Nagao}, {Nakajima}, {Silverman}, {Tanaka}, \& {Wang}}]{ono2017}
{Ono}, Y., {Ouchi}, M., {Harikane}, Y., {et~al.} 2017, ArXiv e-prints,
  arXiv:1704.06004

\bibitem[{{Peng} {et~al.}(2002){Peng}, {Ho}, {Impey}, \& {Rix}}]{peng2002}
{Peng}, C.~Y., {Ho}, L.~C., {Impey}, C.~D., \& {Rix}, H.-W. 2002, \aj, 124, 266

\bibitem[{{Peng} {et~al.}(2010){Peng}, {Ho}, {Impey}, \& {Rix}}]{peng2010}
---. 2010, \aj, 139, 2097

\bibitem[{{Rasappu} {et~al.}(2016){Rasappu}, {Smit}, {Labb{\'e}}, {Bouwens},
  {Stark}, {Ellis}, \& {Oesch}}]{rasappu2016}
{Rasappu}, N., {Smit}, R., {Labb{\'e}}, I., {et~al.} 2016, \mnras, 461, 3886

\bibitem[{{Ribeiro} {et~al.}(2016){Ribeiro}, {Le F{\`e}vre}, {Tasca}, {Lemaux},
  {Cassata}, {Garilli}, {Maccagni}, {Zamorani}, {Zucca}, {Amor{\'{\i}}n},
  {Bardelli}, {Fontana}, {Giavalisco}, {Hathi}, {Koekemoer}, {Pforr}, {Tresse},
  \& {Dunlop}}]{ribeiro2016}
{Ribeiro}, B., {Le F{\`e}vre}, O., {Tasca}, L.~A.~M., {et~al.} 2016, \aap, 593,
  A22

\bibitem[{{Roberts-Borsani} {et~al.}(2016){Roberts-Borsani}, {Bouwens},
  {Oesch}, {Labbe}, {Smit}, {Illingworth}, {van Dokkum}, {Holden}, {Gonzalez},
  {Stefanon}, {Holwerda}, \& {Wilkins}}]{roberts-borsani2016}
{Roberts-Borsani}, G.~W., {Bouwens}, R.~J., {Oesch}, P.~A., {et~al.} 2016,
  \apj, 823, 143

\bibitem[{{Rogers} {et~al.}(2014){Rogers}, {McLure}, {Dunlop}, {Bowler},
  {Curtis-Lake}, {Dayal}, {Faber}, {Ferguson}, {Finkelstein}, {Grogin},
  {Hathi}, {Kocevski}, {Koekemoer}, \& {Kurczynski}}]{rogers2014}
{Rogers}, A.~B., {McLure}, R.~J., {Dunlop}, J.~S., {et~al.} 2014, \mnras, 440,
  3714

\bibitem[{{Ryan} {et~al.}(2011){Ryan}, {Thorman}, {Yan}, {Fan}, {Yan},
  {Mechtley}, {Hathi}, {Cohen}, {Windhorst}, {McCarthy}, \&
  {Wittman}}]{ryan2011}
{Ryan}, R.~E., {Thorman}, P.~A., {Yan}, H., {et~al.} 2011, \apj, 739, 83

\bibitem[{{Sanders} {et~al.}(2007){Sanders}, {Salvato}, {Aussel}, {Ilbert},
  {Scoville}, {Surace}, {Frayer}, {Sheth}, {Helou}, {Brooke}, {Bhattacharya},
  {Yan}, {Kartaltepe}, {Barnes}, {Blain}, {Calzetti}, {Capak}, {Carilli},
  {Carollo}, {Comastri}, {Daddi}, {Ellis}, {Elvis}, {Fall}, {Franceschini},
  {Giavalisco}, {Hasinger}, {Impey}, {Koekemoer}, {Le F{\`e}vre}, {Lilly},
  {Liu}, {McCracken}, {Mobasher}, {Renzini}, {Rich}, {Schinnerer}, {Shopbell},
  {Taniguchi}, {Thompson}, {Urry}, \& {Williams}}]{sanders2007}
{Sanders}, D.~B., {Salvato}, M., {Aussel}, H., {et~al.} 2007, \apjs, 172, 86

\bibitem[{{Schechter}(1976)}]{schechter1976}
{Schechter}, P. 1976, \apj, 203, 297

\bibitem[{{Schenker} {et~al.}(2013){Schenker}, {Robertson}, {Ellis}, {Ono},
  {McLure}, {Dunlop}, {Koekemoer}, {Bowler}, {Ouchi}, {Curtis-Lake}, {Rogers},
  {Schneider}, {Charlot}, {Stark}, {Furlanetto}, \& {Cirasuolo}}]{schenker2013}
{Schenker}, M.~A., {Robertson}, B.~E., {Ellis}, R.~S., {et~al.} 2013, \apj,
  768, 196

\bibitem[{{Schmidt} {et~al.}(2014){Schmidt}, {Treu}, {Trenti}, {Bradley},
  {Kelly}, {Oesch}, {Holwerda}, {Shull}, \& {Stiavelli}}]{schmidt2014}
{Schmidt}, K.~B., {Treu}, T., {Trenti}, M., {et~al.} 2014, \apj, 786, 57

\bibitem[{{Schmidt}(1968)}]{schmidt1968}
{Schmidt}, M. 1968, \apj, 151, 393

\bibitem[{{Scoville} {et~al.}(2007){Scoville}, {Aussel}, {Brusa}, {Capak},
  {Carollo}, {Elvis}, {Giavalisco}, {Guzzo}, {Hasinger}, {Impey}, {Kneib},
  {LeFevre}, {Lilly}, {Mobasher}, {Renzini}, {Rich}, {Sanders}, {Schinnerer},
  {Schminovich}, {Shopbell}, {Taniguchi}, \& {Tyson}}]{scoville2007}
{Scoville}, N., {Aussel}, H., {Brusa}, M., {et~al.} 2007, \apjs, 172, 1

\bibitem[{{S\`ersic}(1968)}]{sersic1968}
{S\`ersic}, J.~L. 1968, {Atlas de galaxias australes}

\bibitem[{{Shibuya} {et~al.}(2015){Shibuya}, {Ouchi}, \&
  {Harikane}}]{shibuya2015}
{Shibuya}, T., {Ouchi}, M., \& {Harikane}, Y. 2015, \apjs, 219, 15

\bibitem[{{Skelton} {et~al.}(2014){Skelton}, {Whitaker}, {Momcheva}, {Brammer},
  {van Dokkum}, {Labb{\'e}}, {Franx}, {van der Wel}, {Bezanson}, {Da Cunha},
  {Fumagalli}, {F{\"o}rster Schreiber}, {Kriek}, {Leja}, {Lundgren}, {Magee},
  {Marchesini}, {Maseda}, {Nelson}, {Oesch}, {Pacifici}, {Patel}, {Price},
  {Rix}, {Tal}, {Wake}, \& {Wuyts}}]{skelton2014}
{Skelton}, R.~E., {Whitaker}, K.~E., {Momcheva}, I.~G., {et~al.} 2014, \apjs,
  214, 24

\bibitem[{{Smit} {et~al.}(2014){Smit}, {Bouwens}, {Labb{\'e}}, {Zheng},
  {Bradley}, {Donahue}, {Lemze}, {Moustakas}, {Umetsu}, {Zitrin}, {Coe},
  {Postman}, {Gonzalez}, {Bartelmann}, {Ben{\'{\i}}tez}, {Broadhurst}, {Ford},
  {Grillo}, {Infante}, {Jimenez-Teja}, {Jouvel}, {Kelson}, {Lahav}, {Maoz},
  {Medezinski}, {Melchior}, {Meneghetti}, {Merten}, {Molino}, {Moustakas},
  {Nonino}, {Rosati}, \& {Seitz}}]{smit2014}
{Smit}, R., {Bouwens}, R.~J., {Labb{\'e}}, I., {et~al.} 2014, \apj, 784, 58

\bibitem[{{Smit} {et~al.}(2015){Smit}, {Bouwens}, {Franx}, {Oesch}, {Ashby},
  {Willner}, {Labb{\'e}}, {Holwerda}, {Fazio}, \& {Huang}}]{smit2015}
{Smit}, R., {Bouwens}, R.~J., {Franx}, M., {et~al.} 2015, \apj, 801, 122

\bibitem[{{Smolcic} {et~al.}(2017){Smolcic}, {Novak}, {Bondi}, {Ciliegi},
  {Mooley}, {Schinnerer}, {Zamorani}, {Navarrete}, {Bourke}, {Karim},
  {Vardoulaki}, {Leslie}, {Delhaize}, {Carilli}, {Myers}, {Baran},
  {Delvecchio}, {Miettinen}, {Banfield}, {Balokovic}, {Bertoldi}, {Capak},
  {Frail}, {Hallinan}, {Hao}, {Herrera Ruiz}, {Horesh}, {Ilbert}, {Intema},
  {Jelic}, {Kloeckner}, {Krpan}, {Kulkarni}, {McCracken}, {Laigle},
  {Middleberg}, {Murphy}, {Sargent}, {Scoville}, \& {Sheth}}]{smolcic2017}
{Smolcic}, V., {Novak}, M., {Bondi}, M., {et~al.} 2017, ArXiv e-prints,
  arXiv:1703.09713

\bibitem[{{Stark}(2016)}]{stark2016}
{Stark}, D.~P. 2016, \araa, 54, 761

\bibitem[{{Stark} {et~al.}(2013){Stark}, {Schenker}, {Ellis}, {Robertson},
  {McLure}, \& {Dunlop}}]{stark2013}
{Stark}, D.~P., {Schenker}, M.~A., {Ellis}, R., {et~al.} 2013, \apj, 763, 129

\bibitem[{{Stark} {et~al.}(2017){Stark}, {Ellis}, {Charlot}, {Chevallard},
  {Tang}, {Belli}, {Zitrin}, {Mainali}, {Gutkin}, {Vidal-Garc{\'{\i}}a},
  {Bouwens}, \& {Oesch}}]{stark2017}
{Stark}, D.~P., {Ellis}, R.~S., {Charlot}, S., {et~al.} 2017, \mnras, 464, 469

\bibitem[{{Stefanon} {et~al.}(2016){Stefanon}, {Bouwens}, {Labb{\'e}},
  {Muzzin}, {Marchesini}, {Oesch}, \& {Gonzalez}}]{stefanon2016}
{Stefanon}, M., {Bouwens}, R.~J., {Labb{\'e}}, I., {et~al.} 2016, ArXiv
  e-prints, arXiv:1611.09354

\bibitem[{{Szalay} {et~al.}(1999){Szalay}, {Connolly}, \&
  {Szokoly}}]{szalay1999}
{Szalay}, A.~S., {Connolly}, A.~J., \& {Szokoly}, G.~P. 1999, \aj, 117, 68

\bibitem[{{Taniguchi} {et~al.}(2007){Taniguchi}, {Scoville}, {Murayama},
  {Sanders}, {Mobasher}, {Aussel}, {Capak}, {Ajiki}, {Miyazaki}, {Komiyama},
  {Shioya}, {Nagao}, {Sasaki}, {Koda}, {Carilli}, {Giavalisco}, {Guzzo},
  {Hasinger}, {Impey}, {LeFevre}, {Lilly}, {Renzini}, {Rich}, {Schinnerer},
  {Shopbell}, {Kaifu}, {Karoji}, {Arimoto}, {Okamura}, \&
  {Ohta}}]{taniguchi2007}
{Taniguchi}, Y., {Scoville}, N., {Murayama}, T., {et~al.} 2007, \apjs, 172, 9

\bibitem[{{Trenti} {et~al.}(2011){Trenti}, {Bradley}, {Stiavelli}, {Oesch},
  {Treu}, {Bouwens}, {Shull}, {MacKenty}, {Carollo}, \&
  {Illingworth}}]{trenti2011}
{Trenti}, M., {Bradley}, L.~D., {Stiavelli}, M., {et~al.} 2011, \apjl, 727, L39

\bibitem[{{van Dokkum} {et~al.}(2011){van Dokkum}, {Brammer}, {Fumagalli},
  {Nelson}, {Franx}, {Rix}, {Kriek}, {Skelton}, {Patel}, {Schmidt}, {Bezanson},
  {Bian}, {da Cunha}, {Erb}, {Fan}, {F{\"o}rster Schreiber}, {Illingworth},
  {Labb{\'e}}, {Lundgren}, {Magee}, {Marchesini}, {McCarthy}, {Muzzin},
  {Quadri}, {Steidel}, {Tal}, {Wake}, {Whitaker}, \&
  {Williams}}]{vandokkum2011}
{van Dokkum}, P.~G., {Brammer}, G., {Fumagalli}, M., {et~al.} 2011, \apjl, 743,
  L15

\bibitem[{{Wilkins} {et~al.}(2016){Wilkins}, {Bouwens}, {Oesch}, {Labb{\'e}},
  {Sargent}, {Caruana}, {Wardlow}, \& {Clay}}]{wilkins2016b}
{Wilkins}, S.~M., {Bouwens}, R.~J., {Oesch}, P.~A., {et~al.} 2016, \mnras, 455,
  659

\bibitem[{{Wilkins} {et~al.}(2011){Wilkins}, {Bunker}, {Stanway}, {Lorenzoni},
  \& {Caruana}}]{wilkins2011}
{Wilkins}, S.~M., {Bunker}, A.~J., {Stanway}, E., {Lorenzoni}, S., \&
  {Caruana}, J. 2011, \mnras, 417, 717

\bibitem[{{Williams} {et~al.}(1996){Williams}, {Blacker}, {Dickinson}, {Dixon},
  {Ferguson}, {Fruchter}, {Giavalisco}, {Gilliland}, {Heyer}, {Katsanis},
  {Levay}, {Lucas}, {McElroy}, {Petro}, {Postman}, {Adorf}, \&
  {Hook}}]{williams1996}
{Williams}, R.~E., {Blacker}, B., {Dickinson}, M., {et~al.} 1996, \aj, 112,
  1335

\bibitem[{{Yan} {et~al.}(2011){Yan}, {Yan}, {Zamojski}, {Windhorst},
  {McCarthy}, {Fan}, {R{\"o}ttgering}, {Koekemoer}, {Robertson}, {Dav{\'e}}, \&
  {Cai}}]{yan2011}
{Yan}, H., {Yan}, L., {Zamojski}, M.~A., {et~al.} 2011, \apjl, 728, L22

\bibitem[{{Zitrin} {et~al.}(2015){Zitrin}, {Labb{\'e}}, {Belli}, {Bouwens},
  {Ellis}, {Roberts-Borsani}, {Stark}, {Oesch}, \& {Smit}}]{zitrin2015}
{Zitrin}, A., {Labb{\'e}}, I., {Belli}, S., {et~al.} 2015, \apjl, 810, L12

\end{thebibliography}

\end{document}